\documentclass[aip,amsmath,amssymb,reprint]{revtex4-1}
\usepackage[pdftex]{graphicx}
\usepackage[english]{babel}

\usepackage[T1]{fontenc}
\usepackage[colorlinks,linkcolor=blue,citecolor=blue,urlcolor=blue]{hyperref}
\hypersetup{pdftitle={Non-Maxwellian electron distribution functions due to self-generated turbulence in collisionless guide-field reconnection},pdfauthor={P. A. Mu\~noz}}

\RequirePackage{snapshot}

\begin{document}
\title[]{Non-Maxwellian electron distribution functions due to self-generated turbulence in collisionless guide-field reconnection}

\author{P. A. Mu\~noz}
\email{munozp@mps.mpg.de}
\affiliation{Max-Planck-Institut f\"ur Sonnensystemforschung, D-37077 G\"ottingen, Germany}
\affiliation{Max-Planck/Princeton Center for Plasma Physics, D-37077 G\"ottingen, Germany}

\author{J. B\"uchner}
\affiliation{Max-Planck-Institut f\"ur Sonnensystemforschung, D-37077 G\"ottingen, Germany}
\affiliation{Max-Planck/Princeton Center for Plasma Physics, D-37077 G\"ottingen, Germany}

\date{\today}

\begin{abstract}
	Non-Maxwellian electron velocity space distribution functions (EVDF) are useful signatures of plasma conditions and non-local consequences of collisionless magnetic reconnection. In the past, EVDFs were obtained mainly for antiparallel  reconnection and under the influence of weak guide-fields in the direction perpendicular to the reconnection plane. EVDFs are, however, not well known, yet, for oblique (or component-) reconnection in case and in dependence on stronger guide-magnetic fields and for the exhaust (outflow) region of reconnection away from the diffusion region.
	In view of the multi-spacecraft Magnetospheric Multiscale Mission (MMS), we derived the non-Maxwellian EVDFs of collisionless magnetic reconnection in dependence on the guide-field strength $b_g$ from small ($b_g\approx0$) to very strong ($b_g=8$) guide-fields, taking into account the feedback of the self-generated turbulence.
	For this sake, we carried out 2.5D fully-kinetic Particle-in-Cell simulations using the ACRONYM code.
	We obtained anisotropic EVDFs and electron beams propagating along the separatrices as well as in the exhaust region of reconnection. The beams are anisotropic with a higher temperature in the direction perpendicular rather than parallel to the local magnetic field. The beams propagate in the direction opposite to the background electrons and cause instabilities.
	We also obtained the guide-field dependence of the relative electron-beam drift speed, threshold and properties of the resulting streaming instabilities including the strongly non-linear saturation of the self-generated plasma turbulence. This turbulence and its non-linear feedback cause non-adiabatic parallel electron acceleration.
	We further obtained the resulting EVDFs due to the non-linear feedback of the saturated self-generated turbulence near the separatrices and in the exhaust region of reconnection in dependence on the guide field strength. We found that the influence of the self-generated plasma turbulence leads well beyond the limits of the quasi-linear approximation, to the creation of  phase space holes and an isotropizing pitch-angle scattering. EVDFs obtained by this way can be used for diagnosing collisionless reconnection by using the multi-spacecraft observations carried out by the MMS mission.\\
	\textit{This article may be downloaded for personal use only. Any other use requires prior permission of the author and the American Institute of Physics.\\
	The following article appeared in  P.A. Mu\~noz and J. B\"uchner, Physics of Plasmas \textbf{23}, 102103 (2016),  and may be found at
	\href{http://dx.doi.org/10.1063/1.4963773}{http://dx.doi.org/10.1063/1.4963773}
	}
\end{abstract}

\maketitle

\section{Introduction}\label{sec:intro}

In astrophysical, space and laboratory plasmas, magnetic reconnection is a crucial mechanism  of conversion of magnetic energy into plasma heating, bulk plasma motion and particle acceleration. In the past, mainly the fluid aspects of this process were analyzed theoretically and numerically, as well as observationally and experimentally.\cite{Buchner2007a,Yamada2010,Zweibel2009,Treumann2013b,Gonzalez2016a} Beyond any fluid description, collisionless magnetic reconnection causes, however, also non-Maxwellian features of the velocity space distribution functions (VDF). Non-Maxwellian VDFs are formed, first of all, by local plasma processes. But they also contain information about the particles' acceleration history and of their interactions with turbulence, providing remote signatures of these processes. Hence, measured VDFs can be used for the investigation of magnetic reconnection in, e.g., the Earth' magnetosphere. While ion-VDFs were already well investigated in the past,  currently electron VDFs (EVDFs) are also becoming increasingly well observed, e.g., by the multi-spacecraft Magnetospheric Multiscale Mission (MMS) with its unprecedent spatial and temporal resolution to measure EVDFs.\cite{Burch2015} Self-consistent kinetic simulations using Vlasov- or Particle-in-Cell (PiC) codes have to be carried out to understand the non-local and non-linear processes forming the EVDFs since hybrid, test particle or fluid approaches do not describe them.

In the past, the formation of EVDFs by reconnection was investigated mainly for the case of antiparallel asymptotic magnetic fields.\cite{Hoshino2001a,Fujimoto2009b,Divin2010, Ng2011, Bessho2014, Shuster2014,Shuster2015, Egedal2015, Cheng2015, Zenitani2016} These investigations obtained typical EVDFs in and close to the diffusion region of reconnection---as long as the feedback reaction via waves and plasma turbulence plays a minor role.\cite{Shuster2015} In this case, therefore, the shape of the EVDFs can be explained in terms of test particle trajectories in appropriately prescribed electromagnetic fields.

Triangular shaped EVDFs are typical  for antiparallel reconnection.\cite{Ng2011} Let $\pm y$ be the direction of the initially antiparallel magnetic fields on both sides of the current sheet (CS) and $x$ the direction across the CS, i.e., of the density gradients in a Harris-sheet equilibrium. Then $x$--$y$ spans the reconnection plane and $z$ is the out-of-reconnection-plane (current) direction. Triangular EVDFs appear in the $v_y$--$v_z$ electron velocity-space plane. They exhibit different striations or filamentary structures called ``beamlets'' (not to be mixed up with ion beamlets due to correlation-modulated scattering\cite{Buchner1991b}).  As it was described for ions in the Earth's magnetotail,\cite{Speiser1965,Buchner1990,Buchner1991b,Buchner1996b} each ``beamlet'' corresponds to groups of particles that succeeded to bounce on their meandering across the CS midplane ($x=0$, along $y$) the same number of times until they are ejected from the CS. The ``beamlet'' with the maximum velocity in the negative $v_z$ direction consists of electrons with the largest number of bounces which, therefore, succeeded to spend the longest time near the maximum reconnection electric field $E_z$. Their direction in the $v_y$--$v_z$ plane reflects the pitch angle between the reconnecting magnetic field component $B_y$ and the out-of-plane component $B_z$ at the location where the electrons lose their gyrotropy. Other structures in the EVDFs are arcs and swirls. They are formed just outside the diffusion region, after the electrons are remagnetized. Ring structures in the outflows jets form cup-like distributions in the 3D velocity space as first shown for the ion-VDFs.\cite{Buchner1996a, Buchner1998,Bessho2014} The temporal evolution of the EVDFs was studied in Ref.~\onlinecite{Shuster2015}.

Flat-top distributions found in the exhaust region\cite{Hoshino2001a,Fujimoto2006a, Egedal2015} are formed due to pitch angle diffusion. This could be due to temperature anisotropies. The latter was shown to be enhanced in case of a small asymptotic electron plasma-$\beta_e$, where $\beta_e$ is the ratio of the electron thermal  pressure to the magnetic pressure. Temperature anisotropies can cause strong parallel (to the magnetic field) electric fields and double layers which additionally accelerate electrons. Flat-top distributions can also be formed by chaotic scattering of electrons that cross the neutral plane where they temporarily lose their gyrotropy.\cite{Buchner1989} The resulting chaotic pitch-angle scattering reduces the anisotropy. Flat-top distributions were observed in the Earth's magnetosphere.\cite{Asano2008,Teste2009}

Similar investigations were used to analyze the formation of EVDFs by reconnection in the presence of finite magnetic guide-fields $B_g = B_z \ne 0$. Note that a finite guide-field strength corresponds to a deviation from the 180$^{\circ}$ angle between the asymptotic fields in antiparallel reconnection. The resulting shear angle of the asymptotic fields is given by
$\phi=2\,{\rm arccos}\left(b_g\middle/\sqrt{1+b_g^2}\right)$ for the normalized guide-field $b_g=B_g/B_{\infty y}$, where $B_{\infty y}$ is the asymptotic reconnection magnetic field strength. Already for small guide fields $0 < b_g  \lesssim0.2$ (corresponding to $165 ^{\circ} \lesssim \phi < 180 ^{\circ}$), the triangular-shaped EVDF structure disappears\cite{Ng2012,Wang2016z} due to the modification of the quasi-adiabatic motion already before the electrons are fully magnetized in the guide field.\cite{Buchner1991a}
For small guide fields, the structures outside the diffusion region (rings, arcs) also disappear, as well as the outflow jets along the separatrices, typical for antiparallel reconnection.\cite{Goldman2011} Note that, since the shape of the EVDFs changes as $b_g^2m_i/m_e$, relatively large mass ratios are needed to investigate the EVDFs  in the diffusion region of finite guide-field reconnection.\cite{Ng2012}

Finite guide-field reconnection applies, e.g.,  to Earth's magnetopause through which density and magnetic field are asymmetrically changing.\cite{Silin2006} Kinetic simulations of reconnection through asymmetric current sheets revealed crescent-shaped EVDFs, which were explained as being due to a normal electrostatic field component generated by the pressure gradients through the CS.\cite{Silin2006,Pritchett2008,Hesse2014,Bessho2016,Chen2016i,Shay2016,Hesse2016,Egedal2016} These predictions are now being verified by MMS spacecraft  observations at the Earth's magnetopause.\cite{Burch2016,Phan2016}
Note, however, that crescent-shaped VDFs are  not a unique signature of asymmetric reconnection. They are obtained also in symmetric 1D CS fields as it was shown by test particle calculations.\cite{Cowley1980,Buchner1996a,Buchner1996b}

Before discussing the EVDFs formed  by stronger guide-field reconnection, let us first clarify one terminology. As it was found earlier, in guide-field reconnection, the plasma density is different at different separatrices:\cite{Kleva1995,Rogers2003,Drake2003,Pritchett2004,Lapenta2010,Markidis2012a,Lapenta2011a,Lapenta2014b} around two anti-symmetrically located separatrices, the plasma density is enhanced, while it drops near the other two separatrices, forming density cavities. We will address these two different near-separatrix regions as ``high-density'' and ``low-density'' separatrices, respectively.

In the case of strong guide fields, the electrons are fully magnetized all the way through the CS. The theory of the linear tearing mode provides an approximate critical threshold $b_{g,{\rm crit}}$ for the transition to a regime of fully magnetized electrons as (see Eq.~(9) in Ref.~\onlinecite{Daughton2005_Karimabadi2005_1}):
\begin{equation}\label{eq:bg_critical}
	b_{g,{\rm crit}}=\sqrt{\rho_i/L}[(T_e/T_i)(m_e/m_i)]^{1/4}=0.26.
\end{equation}
During the non-linear stage of reconnection, the electron magnetization depends in a more complicated way on the plasma parameters. Numerical simulations have shown that this results in a higher threshold $b_{g,{\rm crit}}$.\cite{Le2013} Generally speaking, the magnetization threshold can be determined by calculating the $\kappa$ parameter defined as\cite{Buchner1989,Buchner1991a,Le2013}
\begin{align}\label{eq:kappa}
	\kappa=\sqrt{\Omega_{\rm min}/\omega_{\rm max}}={\rm min}\left(\sqrt{R_B/\rho_{e,{\rm eff}}}\right),
\end{align}
where $\Omega_{\rm min}$ is the minimum frequency of the electron bouncing through the current sheet, $\omega_{\rm max}$ is its maximum gyrofrequency in the minimum field region, $R_B=1/|\hat{b}\cdot\vec{\nabla} \hat{b}|$ is the curvature radius of the magnetic field lines, $\hat{b}=\vec{B}/B$ is the unit vector in the direction of the local magnetic field, $\rho_{e,{\rm eff}}=v_{th,e,{\rm eff}}/\Omega_{ce}=(\sqrt{k_B T_{e,{\rm eff}}/m_e})(m_e/(eB))$ is the electron Larmor radius in the total local magnetic field $B$, and $T_{e,{\rm eff}}=(1/3)(T_{e,xx} + T_{e,yy} + T_{e,zz})$  is the trace of the temperature tensor. The minimum is taken along the magnetic fields lines.
The $\kappa$ parameter was first introduced in Ref.~\onlinecite{Buchner1989} for antiparallel reconnection and generalized in Refs.~\onlinecite{Buchner1991,Buchner1991a} for guide-field reconnection geometries. $\kappa$ depends on the location of the particle's crossing of the central plane of the CS. It dynamically changes in the course of evolving reconnection. $\kappa<1$ causes particles to meander across the CS midplane, $1\lesssim \kappa\lesssim 2.5$ describes weakly magnetized and chaotically scattered particles and $\kappa>2.5$ is characteristic of fully magnetized and gyrotropic particles.

In the strong guide field case, the magnetic moment $\mu=m_ev_{\perp}^2/2B$ is adiabatically conserved and electrons are magnetically trapped around the magnetic field minima via mirror forces. Electrons trapped in the inflowing plasma are advected towards the reconnection X-line together with the frozen-in magnetic flux. Additional trapping can be due to parallel ambipolar electric fields which maintain the plasma quasineutrality.\cite{Egedal2008,Egedal2009,Le2009,Le2010a,Le2013,Egedal2013} The trapping electric field heats the electrons in the parallel direction. These fields can be quantified by a pseudo-potential  (different from the electrostatic potential $\phi$) along the magnetic field $\Phi_{\parallel}$
\begin{equation}\label{eq:phi_par}
	\Phi_{\parallel}(\vec{x}) = \int_{\vec{x}}^{\infty}\vec{E}\cdot\hat{b}\,d\vec{l}.
\end{equation}
The integration is carried out along a magnetic field line ($\vec{l}$) from a position $\vec{x}$ to the ambient plasma. Note that in numerical simulations, the latter is the edge of the simulation box.\cite{Egedal2009,Egedal2013} Where the pseudo-potential $\Phi_{\parallel}$ is maximum, the electron trapping is most efficient. In regions with enhanced $\Phi_{\parallel}$, anisotropic EVDFs are formed by preferential heating in the parallel direction, often detected in magnetospheric observations.\cite{Oieroset2002,Egedal2005,Chen2008a,Scudder2012} These anisotropic EVDFs are formed by a combination of trapped and passing electron populations. The boundary between trapped and passing electrons in the velocity space was given by Eq.~(17) in Ref.~\onlinecite{Egedal2013} as:
\begin{equation}\label{eq:phi_par_boundary}
	{\cal E} -e\Phi_{\parallel} - \mu B_{\infty}=0,
\end{equation}
where the electron kinetic energy is ${\cal E} = {\cal E}_{\parallel} + {\cal E}_{\perp}$, ${\cal E}_{\parallel}=m_ev_{e,\parallel}^2/2$ and ${\cal E}_{\perp}=m_ev_{e,\perp}^2/2$. The trapped population depends only on the magnetic moment ($f=f_{\infty}(\mu B_{\infty})$), while the passing population depends on the acceleration potential (through $f=f_{\infty}({\cal E} -e\Phi_{\parallel})$). $f_{\infty}$ denotes the EVDF at the end of the magnetic field line which is,  in practice, at the edge of the simulation box. This model has been applied successfully to reproduce the anisotropic features of EVDFs observed in the magnetosphere,\cite{Egedal2005,Egedal2010,Egedal2016} in cases where the magnetic moment $\mu$ is conserved.

Note that the acceleration potential  $\Phi_{\parallel}$ also quantifies the anisotropy in the electron pressure tensor by means of an equation of state.\cite{Le2009} The latter can provide an analytical closure of the two-fluid equations by incorporating the kinetic effect of electron trapping. Taking electron trapping into account as a closure for the electron pressure (instead of the commonly used isotropic electron pressure), two-fluid\cite{Ohia2012} and hybrid (ion-kinetic)\cite{Le2016} simulations could reproduce many of the anisotropy-related features obtained by fully kinetic simulations. Such equation-of-state closure provides a smooth transition between the isotropic (for most particles---passing) and the Chew-Goldberger-Low (CGL) double adiabatic equations of state of collisionless plasmas in strong magnetic fields (in which most particles are trapped).\cite{Chew1956}

In contrast to the antiparallel case, where the reconnection electric field accelerates electrons mainly into the exhaust region, in guide-field reconnection electron beams are formed by bipolar electric fields in the cavities of the ``low-density separatrices'' .\cite{Drake2003,Pritchett2004,Swisdak2005,Cattell2005,Pritchett2005,Drake2005,Pritchett2006} Fully 3D simulations have shown that these electron beams  cause streaming instabilities in the cavities of the ``low-density separatrices'' which lead to a strong electrostatic turbulence (see, e.g., Ref.~\onlinecite{Pritchett2004}). In 2.5D configurations, the turbulence is weaker and located away from the CS midplane, where a small projection of the parallel wavenumber onto the reconnection plane exists. The turbulence leads to  thermalization of the beams via pitch-angle scattering forming flat-top distributions.  The formation mechanisms of these VDFs  in dependence on the guide-field strength are, however, still not clear, as well as their shapes in the exhaust region away from the X-line.

While ion beams were found in the reconnection regions of the Earth's magnetotail, in particular, in the Plasma Sheet Boundary Layer (PSBL) (see Refs.~\onlinecite{Ashour-Abdalla1993, Birn2015a} and references therein), electron beams are less often detected. This is likely due to the higher instrumental resolution required.\cite{McFadden1998,Parks2001,Teste2009,Li2015} Perhaps, current MMS observations will provide more information about them as already about field-aligned currents near the PSBL.\cite{Nakamura2016b}
In the PSBL a strong broadband electrostatic noise (BEN) is generated from the lower hybrid frequency $\Omega_{LH}$\cite{Shinohara1998} up to  the electron plasma frequency. The BEN consists of mostly perpendicular (to the magnetic field) propagating waves associated with flat-top EVDFs.\cite{Teste2009} Waves around the lower hybrid frequency are often detected in magnetic reconnection regions of the Earth's magnetotail.\cite{Zhou2011,Zhou2014c}

Electron beams evolve into electron holes via streaming instabilities (see, e.g., Refs.~\onlinecite{Drake2003,Pritchett2004,Elkina2006,Lee2008c,Goldman2008,Lee2011a}).
Electron holes have been observed  in laboratory experiments,\cite{Fox2008} in the Earth's magnetotail\cite{Cattell2005,Khotyaintsev2010} and at the magnetopause.\cite{Retino2006}
Note that electron holes and the associated electrostatic turbulence do not only appear in guide-field reconnection (as originally thought), but they were also found in antiparallel reconnection cases, as soon as large mass ratios $m_i/m_e$ and simulation boxes size are chosen.\cite{Lapenta2011a}

After all these previous investigations, there is still the open question about the formation of EVDFs in dependence on the strength of strong guide magnetic fields, i.e., for shear angles of the reconnecting fields much smaller than $180^{\circ}$. For such component reconnection, the consequences of instabilities, turbulence, particle trapping and scattering for the formation of the EVDFs are not well known, yet. This is true, in particular, for the exhaust region of reconnection. To fill this gap in view of the coming MMS observations, we derive the potentially measurable non-Maxwellian EVDF signatures in dependence on the guide-field strength, taking into account the nonlinear interaction of the electrons with their self-generated turbulence.

This paper is organized as follows. In Sec.~\ref{sec:setup}, we describe the simulation setup and parameters used. In Sec.~\ref{sec:results}, we give a general overview of the EVDFs obtained in the simulations (Sec.~\ref{sec:evdfs}), their distribution in phase space  (Sec.~\ref{sec:phasespace}), the mechanism of beam formation   (Sec.~\ref{sec:beamformation}), the identification of the dominant beam instabilities and the turbulence generated by them (Sec.~\ref{sec:beaminstabilities}). Finally, in Sec.~\ref{sec:conclusion}, we summarize, discuss and give an overview of the expected properties of the EVDFs in dependence on the most probable macroscopic plasma and field parameters, mainly on the guide-field component strength (the opening shear angle) of the reconnecting magnetic fields.

\section{Simulation setup}\label{sec:setup}
We carried out 2.5D (i.e., neglecting variations along the $z$ direction) PiC simulations with the ACRONYM code.\cite{Kilian2012} We initialize the simulations with a  double Harris current sheet equilibrium\cite{Harris1962}
\begin{align}
	\vec{B}(x)&=B_{\infty y}\left[ \tanh\left(\frac{x-L_x/4}{L}\right) -  \tanh\left(\frac{x-3L_x/4}{L}\right) -1 \right]\hat{y} \nonumber \\
&+ B_z\hat{z}. \label{guidefield_b_initial}
\end{align}
Each simulation run is identically initiated except we vary the external guide-field strength $b_g$ (out of the $x$--$y$ reconnection plane, in the $z$ direction).
We normalize the guide-field as $b_g=B_z/B_{\infty y}$, where $B_{\infty y}$ is the asymptotic antiparallel magnetic field (used also for the normalization of the magnetic $\vec{B}$ fields). We simulate CSs for a number of guide fields between $b_g=0$ (antiparallel) and up to $b_g=8$. The direction of the initial density variation  across the CS is $x$. We chose a CS with a halfwidth of $L/d_i=0.5$, a mass ratio $m_i/m_e=100$, a frequency ratio $\omega_{pe}/\Omega_{ce}=4.16$, a temperature ratio of $T_i/T_e=1.0$ and a background plasma density $n_b/n_0=0.2$. $n_0=n_e=n_i$ is the electron/ion plasma number density of the current-carrying population at the center of the CS, $d_{i/e}=c/\omega_{pi/pe}$ is the ion/electron skin depth, $\omega_{pi/pe}$ the ion/electron plasma frequency calculated with the density $n_0$ and $\Omega_{ce}$ is the electron plasma frequency in the asymptotic magnetic field $B_{\infty y}$. The background plasma  has the same temperature and temperature ratio as the main current-carrying population. The previous parameters give an electron thermal speed of $v_{th,e}/c=\sqrt{k_B T_e/m_e}/c=0.12$.
The number of particles per cell (ppc) for the current-carrying  population of electron and ions is $250$ at the center of the CS, while the background is represented by $50$ ppc. The simulation box size in the $x$ and $y$ directions is $L_x\times L_y =(20.94\,d_i\times 12.56\,d_i)$, and the  boundary conditions are periodic in both directions.
The simulation is carried out over $2500\times 1500$ grid points, with a size of each grid-cell $\Delta x=0.7\lambda_{De}$, where $\lambda_{De}$ is the electron Debye length calculated with $n_0$. The Debye length is therefore over-resolved, to provide a sufficient resolution of the electron Larmor radius in the guide field,  $\rho_{e,bg}=v_{th,e}/(b_g\Omega_{ce})$, also in case of the strongest guide magnetic field $b_g=8$. In our case, the ratio $\Delta x/\rho_{e,bg}$ is $0.166$ for $b_g=1$, and it increases linearly with the guide-field. This is to avoid numerical artifacts not related with stability conditions in the regime $\Delta x\gtrsim\rho_{e,bg}$.\cite{Melzani2013} The timestep is chosen as $\Delta t=(1/23.9)\omega_{pe}^{-1}$. The Courant-Friedrichs-Lewy (CFL) condition for light wave propagation is fulfilled at the level $c\Delta t/\Delta x=0.5$.
We initialize the system by a small tearing-like long-wavelength magnetic field perturbation, according to the vector potential $\delta A_z$
\begin{align}\label{perturbation}
	\delta A_z = \delta P B_{\infty y}\frac{L_y}{2\pi}\sin\left(\frac{2\pi \left(y+L_y/4\right)}{L_y}\right)\sin^2\left(\frac{2\pi x}{L_x}\right), 
\end{align}
with an amplitude $\delta P=0.04$. By this way, the developed reconnection stage can be quickly reached with a single large magnetic island and two X-lines centered at $x=\pm L_x/2$ and $y=0$ for each of the two CSs. Further, we depict the results for a zoomed region out of the left CS centered around $x=-L_x/2$.
\section{Results}\label{sec:results}
\subsection{Electron VDFs}\label{sec:evdfs}
Due to the initial perturbation and all the available magnetic flux in the simulation box,  reconnection saturates after $t\gtrsim10\Omega_{ci}^{-1}$. As was found before,\cite{Horiuchi1997, Ricci2004, Pritchett2005} the guide fields delays the onset of reconnection and lower the reconnection rates, mostly due to the Hall effect\cite{Huba2005} of magnetized electrons interacting with the ions.
\begin{figure*}[!ht]
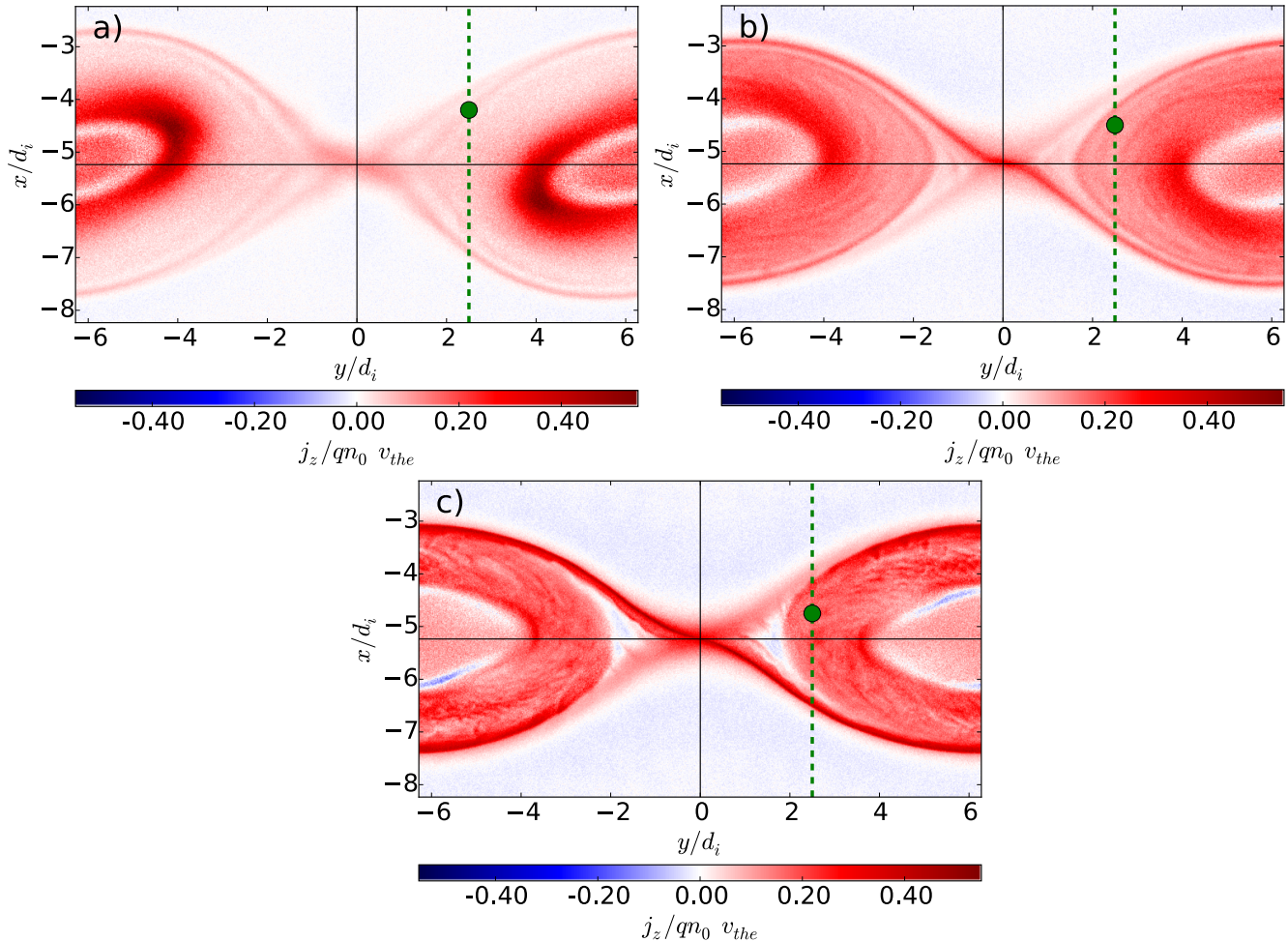
\centering
	\includegraphics[width=0.99\linewidth]{{{./jz_bgs}}}
	\caption{Color-coded contour plot of the current density $j_z$ for different guide fields. (a) $b_g=0.26$ at $t=13\Omega_{ci}^{-1}$, (b) $b_g=1.0$ at $t=14\Omega_{ci}^{-1}$, (c) $b_g=3.0$ at $t=18\Omega_{ci}^{-1}$. The green point is the location used to obtain the EVDF in Fig.~\ref{fig:vdfs_bg}, while the green vertical dashed line is the $x-$cut at constant $y=2.5d_i$ used in Fig.~\ref{fig:phase_space}.\label{fig:jz_bg}}
\end{figure*}

The resulting spatial structure of the out-of-plane current density $j_z$ is shown in Fig.~\ref{fig:jz_bg} for three guide-field cases ($b_g=0.26,\,1.0,\,3.0$). For each case, the corresponding moment of time  is chosen just after the saturation stage, when the opening angle of the exhaust region  and reconnection rates are similar.
Note that a strong plasma turbulence develops in the exhaust region for $b_g=3.0$ (Fig.~\ref{fig:jz_bg}(c)). Turbulence does not develop, however,  at the boundaries in the $y-$direction. This is because the initial current carrying particles accumulate there following the reconnection outflow/exhaust. Thus, this turbulent region resembles the PSBL of the Earth's magnetosphere.
The plasma flow is more laminar in the smaller guide-field $b_g=1$ case (Fig.~\ref{fig:jz_bg}(b)). Turbulence does not develop at all for the smallest guide-field $b_g=0.26$ (Fig.~\ref{fig:jz_bg}(a)).

The three values of $b_g=0.26,\,1.0,\,3.0$ well represent three different regimes of guide-field reconnection. Indeed, the first case $b_g=0.26$ represents the critical guide field for getting magnetized electrons in linear tearing mode theory, obtained by evaluating Eq.~\eqref{eq:bg_critical} for our parameters. For the nonlinear stage of reconnection, such as for the times shown in Fig.~\ref{fig:jz_bg}, we calculated the more accurate magnetization parameter $\kappa(\vec{x})$ defined by Eq.~\eqref{eq:kappa} for the magnetic field lines in the reconnection exhaust (plots not shown here). For the case of $b_g=0.26$, $\kappa(\vec{x})$  increases in time, from values  $\kappa\gtrsim 1$ for $t=6\Omega_{ci}^{-1}$ to around  $\kappa\sim 2-3$ after $t=14\Omega_{ci}^{-1}$.
In this limit, the electrons are quasi-adiabatically scattered and their orbits are chaotic.\cite{Buchner1989,Buchner1991,Buchner1991a,Le2013} For smaller guide fields such as $b_g=0.13$, the initially meandering in the exhaust electrons (for $\kappa < 1$)  change to pitch-angle scattered gyrotropic ($\kappa\gtrsim 1$) motion at later times. For guide fields larger than $b_g=0.53$, the electrons are fully magnetized and gyrotropic from the very beginning with, typically, $\kappa\gtrsim 2.5$. Therefore, for $b_g=1$ and $b_g=3$ shown in Figs.~\ref{fig:jz_bg}(b) and~\ref{fig:jz_bg}(c), the electrons are fully magnetized and gyrotropic everywhere, and their magnetic moment $\mu$ is conserved to a large degree. Note that because the transition between different regimes of reconnection depends on $(B_g/B_{\infty y})\sqrt{m_i/m_e}$ as found by Ref.~\onlinecite{Le2013}, the same behaviour can be expected for smaller guide fields if higher (more realistic) mass ratios $m_i/m_e$ are used.

\begin{figure*}[!ht]
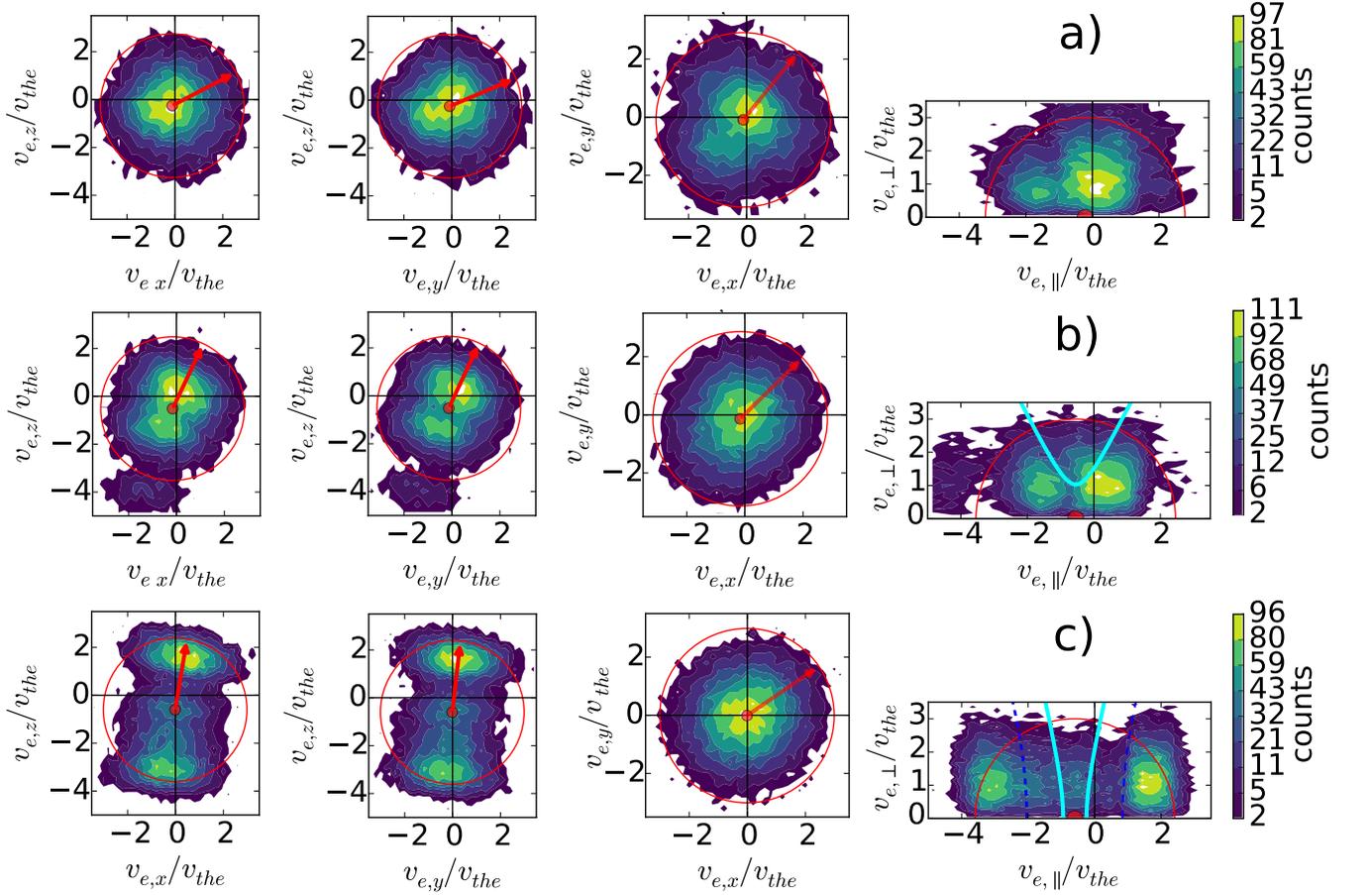
 \centering
	\includegraphics[width=0.99\linewidth]{{{./detailfullvelspace_bgs3}}}
	\caption{EVDFs for guide fields $b_g=0.26$ (row a), $b_g=1.0$ (row b) and $b_g=3.0$ (row c), obtained at the locations in the exhaust region shown in Fig.~\ref{fig:jz_bg} (at the same times). For each case, three different combinations of velocity space components in addition to a distribution with the perpendicular and parallel velocity components are shown. The EVDFs were calculated in a square region of size $(0.1\times0.2)d_i$. The red point indicates the mean drift speed, the red empty circle with radius of $3v_{th,e}$ indicates the thermal spread of an isotropic Maxwellian EVDF (centered in the mean drift speed), and the red arrow indicates the direction of the local magnetic field. The cyan lines in the $v_{\parallel}$--$v_{\perp}$ velocity space represent the boundaries between trapped and passing electron populations given by Eq.~\eqref{eq:phi_par_boundary} and calculated with local parameters. The blue dashed lines for the case $b_g=3$ represent that boundary using a hypothetically higher acceleration potential value $e\Phi_{\parallel}/k_BT_e \sim 1.0$. \label{fig:vdfs_bg}}
\end{figure*}

Three characteristic EVDFs for the three guide field cases of Fig.~\ref{fig:jz_bg} are shown in Fig.~\ref{fig:vdfs_bg} for a point between the ``low-density separatrix'' and the exhaust region (shown in Fig.~\ref{fig:jz_bg}).
The distributions in the plane $v_x$--$v_z$ and $v_y$--$v_z$ demonstrate that the EVDFs deviate more from a Maxwellian distribution for stronger guide fields. The reason for this deviation can be found by analyzing the  turbulence developed inside the exhaust region.

As one can see in Fig.~\ref{fig:vdfs_bg}(c), a structure that resembles a double beam forms in the strong guide field case $b_g=3.0$. The faster beam propagates in the direction opposite to the local magnetic field  (depicted as an red arrow), mainly in the $-z$ direction, in agreement with previous works (discussed in Sec.~\ref{sec:intro}). The mean drift speed of the faster beam moving away from the X-line can be estimated as $\vec{V}_d/v_{th,e}=-0.36\hat{x}-0.28\hat{y}-3.08\hat{z}$. The beam in the opposite direction propagates towards the X-line with $\vec{V}_d/v_{th,e}=0.34\hat{x}+0.23\hat{y}+1.59\hat{z}$.
The counterstreaming beams may cause instabilities (the dip between the beams evolves into a phase space hole) as well as the anisotropy of the beams with $T_{e,\perp}>T_{e,\parallel}$. Specifically, the faster beam has $v_{th,e\parallel}/v_{th,e}=0.58$ and $v_{th,e\perp}/v_{th,e}=1.01$, while the slower beam has $v_{th,e\parallel}/v_{th,e}=0.49$ and $v_{th,e\perp}/v_{th,e}=1.01$. In agreement with previous investigations, the EVDFs are slightly non-gyrotropic, to a larger extent near the ``low-density separatrix'' (results not shown here).\cite{Wendel2016} This non-gyrotropy was already detected  in magnetospheric observations.\cite{Chen2008a,Scudder2012,Norgren2016} The decay channel of the temperature anisotropy is likely through the electron whistler instability, possible even in the low-plasma-$\beta_e$ conditions present here, generating parallel propagating waves (see Sec.~7.3.2-3 in Ref.~\onlinecite{Gary1993}). These whistler waves due to temperature anisotropy were also detected in magnetopause observations.\cite{Graham2016}

In the moderately strong $b_g=1.0$ case (Fig.~\ref{fig:vdfs_bg}(b)), the distribution is anisotropic along the magnetic field direction, with $T_{e,\parallel}$>$T_{e,\perp}$. There is no clear double beam structure, only a barely visible double peaked Maxwellian with components much closer each other than for the $b_g=3.0$ case (compare especially the EVDFs in the plane  $v_{\parallel}$--$v_{\perp}$).
For the smallest guide-field $b_g=0.26$ (Fig.~\ref{fig:vdfs_bg}(a)), the EVDF is very close to an isotropic Maxwellian, not developing a tail or plateau in the magnetic field direction. This is because the electrons spent less time being accelerated near the X-line by the reconnection electric field $E_z$ than for stronger guide-field cases. It is still possible to distinguish, however, a small ``bump'' oppositely directed to the local magnetic field, especially in the plane  $v_{\parallel}$--$v_{\perp}$. The main consequence of these EVDFs structures for both $b_g=1.0$ and $b_g=0.26$ cases is that both of them do not become unstable and do not generate any significant turbulence.

\begin{figure*}[!ht]
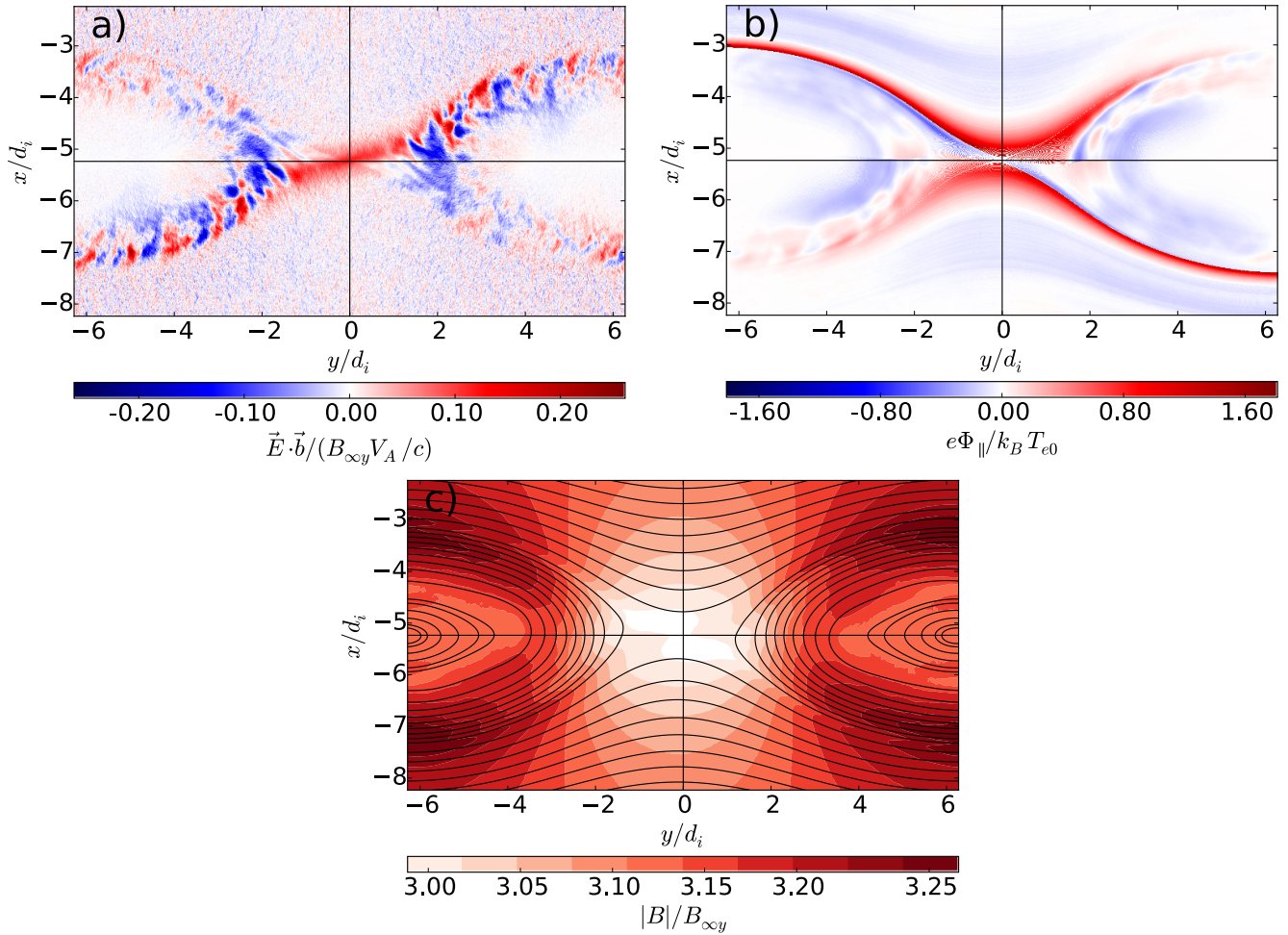
\centering
	\includegraphics[width=0.99\linewidth]{{{./trapping_bg3}}}
	\caption{Color-coded contour plots of quantities related with electron trapping for the case $b_g=3.0$ at $t=18\Omega_{ci}^{-1}$. (a) Parallel electric field $E_{\parallel}=\vec{E}\cdot\vec{b}$. (b) Acceleration potential $\Phi_{\parallel}$ (Eq.~\eqref{eq:phi_par}). (c) total magnetic field $|B|/B_{\infty y}$. Black lines are the  magnetic field lines. \label{fig:trapping_bg3}}
\end{figure*}

The EVDFs in the strongly magnetized ($\kappa>2.5$) cases $b_g=3.0$ and $b_g=1.0$ can be partially explained due to the electron trapping (see discussion in Sec.~\ref{sec:intro}).
Two key quantities control the trapping. One of them is the acceleration potential $\Phi_{\parallel}$ obtained by integrating the parallel electric field $E_{\parallel}=\vec{E}\cdot\vec{b}$ according to Eq.~\eqref{eq:phi_par}.
Fig.~\ref{fig:trapping_bg3}(a) shows that the parallel electric field $E_{\parallel}$ is patchy and bipolar in the guide field case $b_g=3.0$.  It is enhanced in the ``low density separatrix'' (see Sec.~\ref{sec:intro}), although also present in both exhaust region and in the ``high density separatrix'', contributing to the turbulence (see Sec.~\ref{sec:beaminstabilities})  and the formation of non-gyrotropic EVDFs.\cite{Wendel2016} $E_{\parallel}$ is positive near the X-line. Fig.~\ref{fig:trapping_bg3}(b) shows that the acceleration potential $\Phi_{\parallel}$ reaches its maximum values  in a narrow region close to the ``high density separatrix'', and in a wider region near the X-line correlated with the location of positive values of $E_{\parallel}$.
In contrast to the predictions of the trapping model,\cite{Egedal2009,Egedal2013} $\Phi_{\parallel}$ does not correlate well with the regions of enhanced density $n_e$ or field aligned electron temperature anisotropy $T_{e,\parallel}/T_{e,\perp}$ (plots not shown here). Note that the the electric field $E_{\parallel}$ integrated along the full magnetic field lines, $\int_{-\infty}^{\infty} \, E_{\parallel} \, d\vec{l}$, is not negligible small (plot not shown here). In fact, $e\int_{-\infty}^{\infty}E_{\parallel}d\vec{l}/(k_BT_e)\gtrsim 1$ in the region close to the separatrices. As a result, the magnitude of $\Phi_{\parallel}$ depends on the direction of the integration along the magnetic field lines, i.e., $\int_{\vec{x}}^{\infty}E_{\parallel}\,d\vec{l}\neq \int_{\vec{x}}^{-\infty}E_{\parallel}\,d\vec{l}$. This is,  perhaps, due to the finite values of $E_{\parallel}$ at the boundaries (see discussion about the assumptions in Ref.~\onlinecite{Egedal2009} and Eq.~(11) of Ref.~\onlinecite{Egedal2013}).
Note that even the maximum acceleration potential $e\Phi_{\parallel}$ is just barely larger than the thermal energy $k_BT_e$, with even smaller values inside the exhaust region. They may become larger for larger simulation domains and mass ratios $m_i/m_e$.

As the total magnetic field is minimum, the acceleration potential  $\Phi_{\parallel}$ is mainly concentrated in an elliptical region near the X-line (see Fig.~\ref{fig:trapping_bg3}(c)). Because of the relatively small values of  $\Phi_{\parallel}$, the electron trapping due to the electric and magnetic fields is dominated by the latter, i.e., by the mirror force arising from the conservation of the magnetic moment (i.e., perpendicular adiabatic electron dynamics).
We checked the relative contribution of electric and magnetic trapping to the EVDFs shown in Fig.~\ref{fig:vdfs_bg}(c) by calculating the theoretical boundary between trapped and passing population given by Eq.~\eqref{eq:phi_par_boundary}. For this, we use the local parameters at the location where the EVDF was taken, $|B|/B_{\infty}=3.07/3.24$ and $e\Phi_{\parallel}/k_BT_e = 0.06$ (see Figs.~\ref{fig:trapping_bg3}(b) and ~\ref{fig:trapping_bg3}(c)).  The results are depicted in the $v_{\parallel}$--$v_{\perp}$ plot by cyan colored lines (see Fig.~\ref{fig:vdfs_bg}(c)). Here, we took into account the net parallel bulk speed calculated for the entire EVDF. Our result does not agree neither with the anisotropy nor with the beams we obtained: the trapped population (low values of the EVDF $f$ in the direction $v_{\perp}$) it is not in between those lines, while the passing population (high values of EVDF $f$ in the direction $v_{\perp}$) does not match with the beams.
Instead, in order to reproduce the boundary between the trapped/passing populations, a potential on the order of $e\Phi_{\parallel}/k_BT_e \sim 1.0$  would be necessary, as shown by dashed blue lines in Fig.~\ref{fig:vdfs_bg}(c).  Such potential is not available in the exhaust region but only along the non-reconnected, open magnetic fields lines in the inflow region outside the separatrices.

The previous observations allows us to conclude that the trapping model leading to Eq.~\eqref{eq:phi_par_boundary} cannot predict accurately the EVDF features in the case $b_g=3$ at the separatrices. There are at least four reasons invalidating partially this model for this case. First, the features of the acceleration potential $\Phi_{\parallel}$ mentioned above not taken into account in that model. Second, we checked that the off-diagonal terms of the electron pressure tensor significantly contribute to the electric field close to the ``low-density separatrix'', in agreement with previous works\cite{Wendel2016} (results not shown here). This resulting electron non-gyrotropy violates the assumptions of the trapping model.\cite{Le2010} Third, we also checked that charge quasineutrality, one of the main reasons for the presence of the electric field that can trap electrons,\cite{Egedal2008} is not well satisfied near the ``low-density separatrix'' and in some regions in the exhaust of reconnection: deviations from quasineutrality ($(n_i-n_e)/n_e$) can reach values as high as $\sim70\%$ (plots not shown here). This is also a consequence of using a relatively small mass ratio $m_i/m_e=100$: electrons can more efficiently screen electric fields generated due to the slower motion of the ions if their thermal speed $v_{th,e}$ is much faster than the ion one $v_{th,i}$. And fourth, the typical electron drift speeds (see red dot in the fourth panel of Fig.~\ref{fig:vdfs_bg}(c)) are comparable to the electron thermal speed $v_{th,e}$. This violates the assumption that $v_{th,e}$ should be much greater than any other flow speed in the system, implying  that the possibly trapped electrons cannot carry parallel currents.\cite{Egedal2008,Le2009, Le2010} The required high values of the electron thermal speed can also be written in the form $v_{th,e}\gg V_A$ (with $V_A$ the Alfv\'en speed), necessary to have an electron transit and bounce motion periods much smaller than the time scale variation of $\Phi_{\parallel}$.\cite{Le2010} This form of the assumption is also invalid in our case, because the local values of $v_{th,e}$ are actually comparable to the local Alfv\'en speed in a significant fraction of the simulation box (plot not shown here). Note that $v_{th,e}/V_A$ scales as $\sqrt{m_i/m_e}$, and therefore, the separation of scales between both speeds will be larger, and the trapping model more accurate, for more realistic (higher) mass ratios than the value $m_i/m_e=100$ used here.\cite{Le2013} Note that the same result is obtained  by varying either the ion or electron mass. In our simulation setup, it is possible to get higher mass ratios by choosing higher ion masses, but always keeping a fixed electron mass, and therefore a fixed electron thermal speed $v_{th,e}$.

For this case $b_g=3$, the violation of the previous assumptions leads to a non-adiabatic parallel electron dynamics. As a consequence, some electrons originally belonging to the passing population in the right beam can be transferred to the trapped population (see, e.g., Fig.~\ref{fig:trajectories}(c)), and vice-versa (see, e.g., Fig.~\ref{fig:trajectories}(a)), even before the velocity space instabilities saturates (see Sec.~\ref{sec:beaminstabilities}). This also means that not only electrons with initially small parallel energies (i.e., small ${\cal E}_{\parallel,\infty}$) can become trapped, in contrast to the assumptions of the electron trapping model.\cite{Egedal2013} Thus, the parallel and perpendicular electron beam temperatures $T_{e,\parallel}$ and $T_{e,\perp}$ readjust with contributions from both populations during the evolution of the system in the exhaust and separatrices of reconnection, a process not fully captured by the trapping model.

On the other hand, as long as the electrons behave adiabatically in the parallel direction, the trapping model is quite accurate. This applies, e.g., to the case $b_g=1.0$ where there is lack of turbulence in the exhaust region. The corresponding curve $v_{\parallel}$--$v_{\perp}$ predicted by Eq.~\eqref{eq:phi_par_boundary} for the case $b_g=1.0$ is depicted by a cyan colored line in Fig.~\ref{fig:vdfs_bg}(b), in good agreement with the two peaks in the EVDF. We found this by using the local values $|B|/B_{\infty}=1.25/1.55$ and $e\Phi_{\parallel}/k_BT_e=-0.26$ (plots not shown here).

This dependence of the transition to turbulence on the guide field strength can be understood based on the non-adiabatic electron dynamics in the case of small plasma-$\beta_{e,\infty}$, where $\beta_{e,\infty}=2\mu_0n_ek_BT_{e}/B^2$ with all quantities calculated far away in the ambient plasma, at the boundary of the simulation domain.
As it was found before, $\beta_{e,\infty}$ is the key parameter controlling the electron energization along the separatrices and the formation of double layers in antiparallel\cite{Egedal2012,Egedal2015} and guide-field reconnection\cite{Le2013} (see also, for magnetospheric observations, Ref.~\onlinecite{Egedal2010}). Those authors found that the acceleration potential $\Phi_{\parallel}$ scales as $\beta_{e,\infty}^{-1/2}$, and, consequently, also the electron energization and formation of double layers.  In our simulations, $\beta_{e\infty}=0.1$, calculated with $B_{\infty y}$, corresponds to the threshold $\beta_{e\infty,\text{threshold}}=\sqrt{m_e/m_i}=0.1$ as it was derived for the antiparallel case.\cite{Egedal2015} Even though in our case the acceleration potential does not seem to play the essential role for the non-adiabatic electron motion, a threshold in $\beta_{e\infty}$ appeared to be critical for the development of turbulence.  Including the guide field in the calculation of the plasma-$\beta_e$ (i.e., using the total magnetic field at the infinity $\sqrt{B_{\infty y}^2+B_g^2}$), one obtains $\beta_{e\infty}=0.01$ for  $b_g=3$. For this guide field, strong non-Maxwellian features and turbulence develop.  On the other hand, the case $b_g=1$ (practically laminar) has $\beta_{e\infty}=0.05$ while  $b_g=2$ (marginally turbulent) has $\beta_{e\infty}=0.02$.  This indicates that the transition between turbulent/non-turbulent (or non-adiabatic/adiabatic parallel electron) behavior  happens for a small enough plasma-$\beta_{e\infty,\text{critical}}\approx 0.02$, obtained for the total magnetic field at the infinity. Although it is smaller than the value derived in Ref.~\onlinecite{Egedal2015} ($\beta_{e\infty,\text{threshold}}=\sqrt{m_e/m_i}=0.1$), the existence of a threshold depending on $\beta_{e\infty}$ is a strong indication that the non-adiabatic parallel electron behavior might be responsible in part for the formation of non-Maxwellian EVDFs also in our case.

Note that since the threshold for the non-adiabatic parallel electron behavior depends on the mass ratio, we expect that for higher mass ratios $m_i/m_e$, these effects would diminish, since they would require even smaller  $\beta_{e\infty}$ and, consequently, much stronger guide fields or higher values of the asymptotic plasma-$\beta_{e\infty}$. Equivalently, for more realistic mass ratios, the parallel electron dynamics would be more adiabatic and, therefore, the electron trapping model more accurate.

\subsection{Phase space structures}\label{sec:phasespace}
\begin{figure*}[!ht]\centering
	\includegraphics[width=0.99\linewidth]{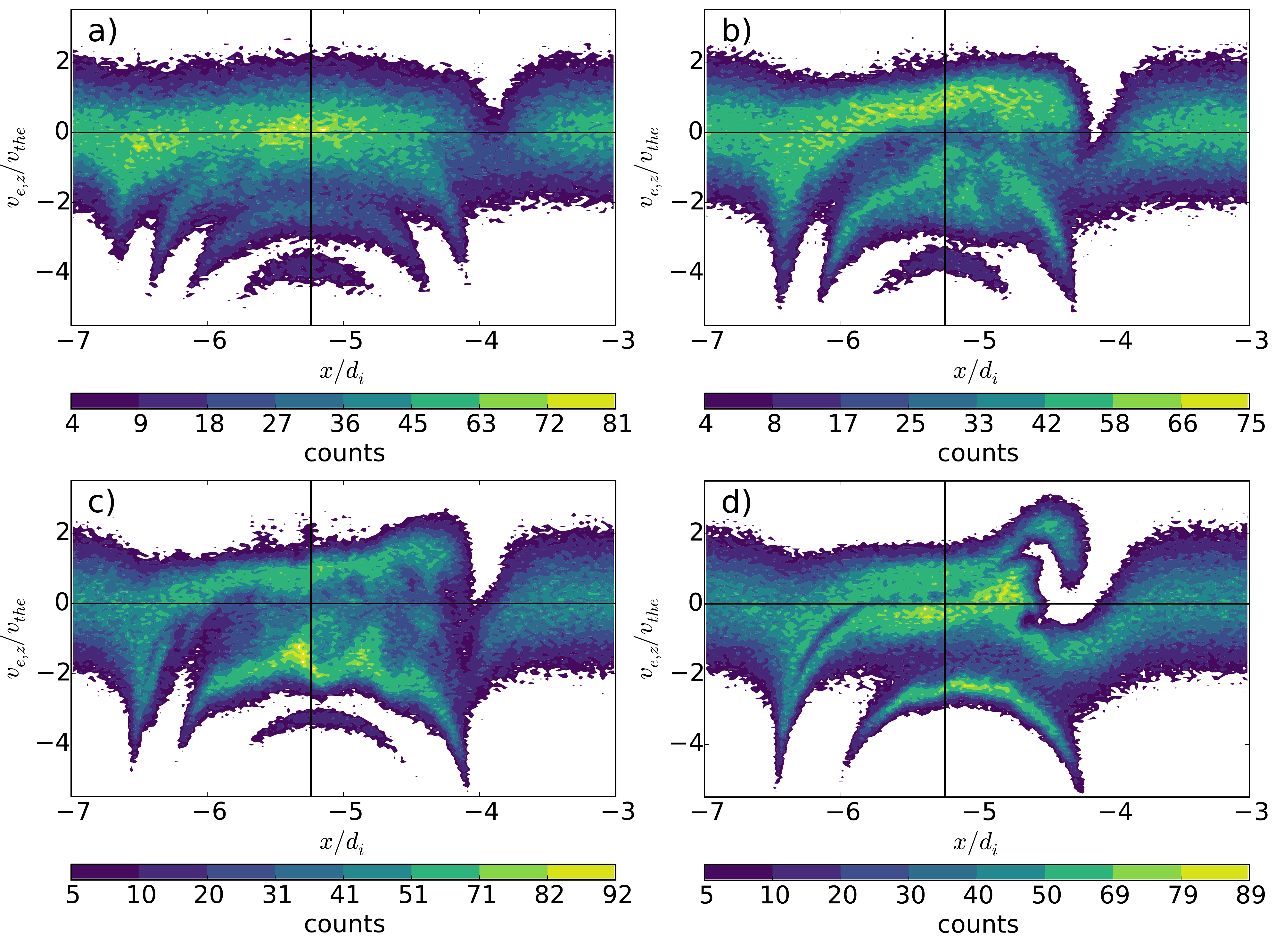}
	\caption{Crescent-shaped electron distribution functions in the phase space $x$--$v_z$ cut at the $y=2.5d_i$ location shown in Fig.~\ref{fig:jz_bg} for different guide fields strengths. (a) $b_g=1$ at $t=14\omega_{ci}^{-1}$. (b) $b_g=2$ at $t=16\omega_{ci}^{-1}$. (c) $b_g=3$ at $t=18\omega_{ci}^{-1}$. (d) $b_g=5$ at $t=20\omega_{ci}^{-1}$. The vertical black line represents the CS center.\label{fig:phase_space}}
\end{figure*}

Electron beams, evolving later into structures similar to phase space holes, typically develop in the whole exhaust region between the separatrices for the case $b_g=3$. This can be seen in phase space plots $x$--$v_z$ shown  in Fig.~\ref{fig:phase_space} for different guide fields, obtained as an $x-$cut at $y=2.5d_i$ (in the exhaust region, see Fig.~\ref{fig:jz_bg}).
In all the cases, the ``high-density separatrix'' is represented as the leftmost peak with $v_z<0$. A density cavity forms around the ``low-density separatrix''  (rightmost dip with $v_z\gtrsim0$). The differences in the separatrices densities are more expressed, the larger the guide fields strength is. The reason is that strong guide fields enhance acceleration of the plasma flows. These plasma flows are then scattered along the separatrices (see, e.g., Chp.~8.4.1 of Ref.~\onlinecite{Gonzalez2016a}).
For higher guide-field strengths,  bipolar electrostatic fields occur  along the ``low-density separatrix'' (see Fig.~\ref{fig:trapping_bg3}(a)).
The maximum speed in $v_z$ of the electrons in the ``high-density separatrix'' is enhanced, while the width of their distribution in $x$ becomes narrower (strong density gradient).

Note that for the guide-field case $b_g=3$ (Fig.~\ref{fig:phase_space}(c)), the distribution of the plasma population in the ``high-density separatrix'' constitutes one arc of a crescent-like (parabolic) shape in the phase space cut $x$--$v_z$ (at $x\sim-6.7d_i$). The other end of the arc is next to the ``low-density separatrix'' (at $x\sim-4.1d_i$). Note that these arcs are asymmetrical or tilted, suggesting a relation with some simulations of beam-driven lower hybrid instability,\cite{McMillan2007} that explained those phase space structures as a result of the wave steepening associated with this instability (see discussion in Sec.~\ref{sec:beaminstabilities}).
In most of the exhaust region, the mean drift speed of the electrons forming the upper arc or crescent-shaped population is $\vec{V}_d\sim +1v_{th,e}\hat{z}$, increasing towards the ``low-density separatrix''. There is a second internal crescent-shaped distribution, with average drift speed of  $\vec{V}_d\approx -3v_{th,e}\hat{z}$ inside the exhaust region and boundaries between $-6.2d_i\lesssim x\lesssim-4.1d_i$. Therefore, throughout this region, the two crescent-shaped populations form the two beams (drifting away mostly in the $z-$direction) similar to those seen in Fig.~\ref{fig:vdfs_bg}(c) (taken at $x=-4.7d_i$). The structures in between resemble phase space holes of size $w\sim0.3d_i$ wide. For smaller guide fields (e.g., $b_g=2$), the dip between the two beams is reduced. This is seen in Fig.~\ref{fig:phase_space}(b) as a reduction in the distance between the two crescent-shaped populations close to the CS center. For $b_g\lesssim1$ (Fig.~\ref{fig:phase_space}(a)), the two crescent-shaped populations completely merge in the exhaust region, becoming completely stable.

The different crescent-shaped populations in the phase space cut $x$--$v_z$ correspond to electrons which were accelerated  at different times and distances from the X-line, as predicted theoretically by Ref.~\onlinecite{Buchner1991} and simulated by Ref.~\onlinecite{Buchner1996b}. Electrons accelerated earlier and closer to the X-line obtain a higher speed in the $-v_z$ direction, being pushed inwards to the exhaust region at an earlier time. Those high energetic electrons are located in the boundaries of the crescent-shaped distribution in the phase space plane $x$--$v_z$ (closer to the separatrices), where they are accelerated first. Near the midplane they are decelerated instead, resulting in the concave shape of the crescent in the phase space cut $x$--$v_z$.
This process is due to the conservation of the magnetic moment $\mu=m_ev_{e,\perp}^{2}/2B$: since the  in-plane $B_{\perp}$ magnetic field is weaker (stronger) near the CS midplane (separatrices), the electron $v_z$ drift speed has to be smaller (larger) there. On the other hand, because the out-of-plane magnetic field $B_z$  is stronger (weaker) near the CS midplane (separatrices), this invariance also implies that the electron in-plane $v_{e,\perp}$ drift speeds are larger (smaller) there (results not shown here).
The smaller the guide-field, the more crescent shaped distributions are formed (see Fig.~\ref{fig:phase_space}(a)). This is because electrons experience a greater acceleration near the  X-line with maximum value of the reconnection electric field, but spend shorter time there compared to the case of stronger guide fields.
As a result,  electrons are trapped in magnetic field lines closer to each other than in stronger guide fields.
\subsection{Beam formation}\label{sec:beamformation}
In order to reconstruct the acceleration of the electron beams, we selected all the electrons ($\approx 16400$) belonging to the EVDF shown in Fig.~\ref{fig:vdfs_bg}(c) for the case $b_g=3$ at $t=16.1\Omega_{ci}^{-1}$. Then, we re-run the simulation outputting at a higher cadence  the particle data, in order to trace the trajectories of the beam particles.
\begin{figure*}[!ht]
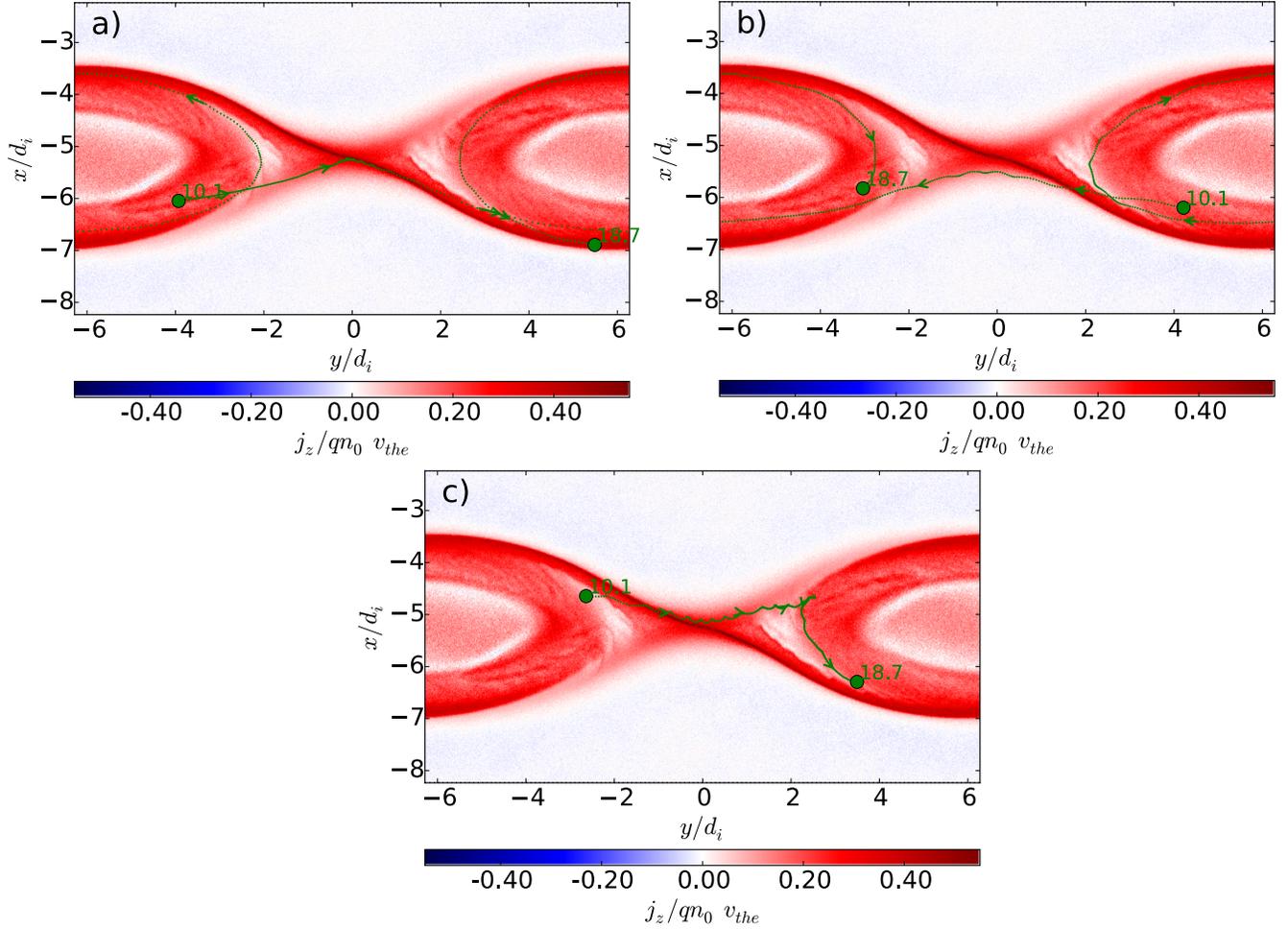
\centering
	\includegraphics[width=0.99\linewidth]{{{./trajectories_bg3}}}
	\caption{Trajectories of selected electrons forming the beam distributions shown in Fig.~\ref{fig:vdfs_bg}(c) for the case $b_g=3$. The contour plot shows the current density $j_z$ at $t=16\Omega_{ci}^{-1}$. The trajectories are shown for a time period between $10.1<t\Omega_{ci}<18.7$. The start and end points are labeled with a larger green point and the corresponding time. The direction of motion is indicated with arrows. \label{fig:trajectories}}
\end{figure*}
\begin{figure*}[!ht]
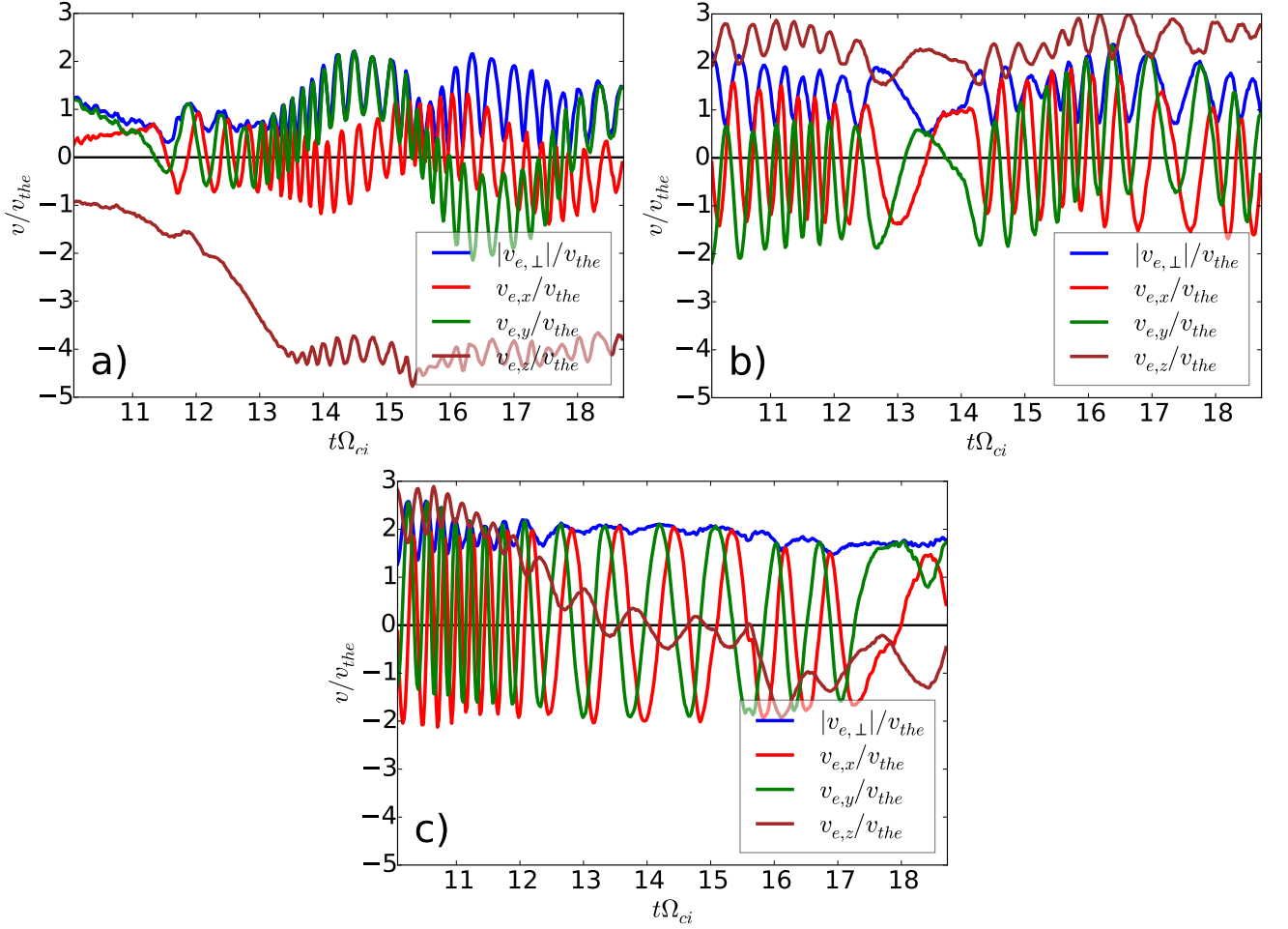
\centering
	\includegraphics[width=0.99\linewidth]{{{./time_histories_bg3}}}
	\caption{Time history of the corresponding velocity components of each particle trajectory of Fig.~\ref{fig:trajectories}  for the case $b_g=3$. \label{fig:time_histories}}
\end{figure*}
We found that most of the electrons in the (faster) $-v_z$ beam (population I) entered the vicinity of the X-line from a ``low-density separatrix''  (Fig.~\ref{fig:trajectories}(a)) and become accelerated by the reconnection electric field.
From there, they are expelled into the direction of  the ``high-density separatrix''.
Obviously, the electrons spend a longer time in the acceleration region the larger the guide-field is.\cite{Swisdak2005} For $b_g=3$, this is visualized as a  strong drop in $v_z$ between $10<t\Omega_{ci}<13.5$ (see Fig.~\ref{fig:time_histories}(a)).
The longer acceleration time is, however, compensated by the smaller reconnection electric field these electron see. As a result, the overall acceleration and final drift speed in the beam are not significantly larger than for lower guide-field cases. On the other hand, after the electrons enter the exhaust region, their large $-v_z$ speed is associated with an in-plane ($x$--$y$) curved drift motion along the magnetic field lines. This drift  is directed outwards from the X-line in the region between the ``high-density separatrix'' and the CS-midplane, and inwards to the X-line in the region between the  ``low-density separatrix'' and the CS-midplane.
As a result, the beam EVDF of this population has a mean drift speed  with negative $-v_x$ and $-v_y$ components for the location shown in Fig.~\ref{fig:jz_bg}(c) (see discussion of  Fig.~\ref{fig:vdfs_bg}(c)).

A typical electron trajectory with initially positive velocity $+v_z$ (population II) is shown in Fig.~\ref{fig:trajectories}(b) (for the EVDF shown in the Fig.~\ref{fig:vdfs_bg}(c)). Their $v_z$ speed does not change significantly (Fig.~\ref{fig:time_histories}(b)): they approach the X-line from the ``high-density separatrix'', but do not encounter it as close as the ones entering via the ``low-density separatrix''. Most of them belonged originally to the background population. Inside of the exhaust region, they have the opposite drift motion to those of the beam electrons with $-v_z$, i.e., directed inwards to the X-line in the region between the ``high-density separatrix'' and the CS midplane, and outwards from the X-line in the region between the  ``low-density separatrix'' and the CS-midplane.
As a result, and different from the previous case, their beam EVDF has a mean drift speed  with positive $v_x$ and $v_y$ components for the location shown in Fig.~\ref{fig:jz_bg}(c).

The counterstreaming beam electrons cause an instability whose main sources of free energy are the in-plane projection  of the relative  drift speed between them and the relative speed of each one with respect to the ions (practically at $\vec{v}_{\parallel}=0$ compared with the electrons). The beam instability vanishes if the guide fields strength exceeds significantly $b_g=5$. Even though the beam in the negative $-v_z$ direction is slightly faster in this limit (due to the extra time spent in the acceleration region close to the X-line), its velocity projection in the $x$--$y$ plane is smaller due to the stronger magnetization. For small guide fields $b_g\leq1$, on the other hand, there is no formation of electrons beams but rather a barely visible double peaked Maxwellian, and therefore, no significant source of free energy for instabilities (see also discussion of Fig.~\ref{fig:phase_space}).
Finally, there is a third electron population (population III) illustrated by the trajectory shown in Fig.~\ref{fig:trajectories}(c). These particles come from one of the ``high-density separatrices''. Their $v_z$ speed component is positive, being decelerated near the X-line, and then ejected into the ``low-density separatrix''.
Inside the exhaust region, the electrons of this population are reflected, changing their curved drift motion  (Fig.~\ref{fig:time_histories}(c)) from the characteristic direction of population II (with $+v_z$) to those of the population I (with $-v_z$). This can be better understood from the theory of electron trapping in magnetic and electric fields discussed in the Sec.~\ref{sec:intro} and Sec.~\ref{sec:evdfs}. Indeed, the reflection of a typical electron takes place in the region near the X-line and the exhaust region where the magnetic field is a minimum (Fig.~\ref{fig:trapping_bg3}(c)) and the acceleration potential $\Phi_{\parallel}$ is significant (Fig.~\ref{fig:trapping_bg3}(b)). But due to the small values of $\Phi_{\parallel}$, magnetic trapping should be predominant in the exhaust region. This also explains the small number of bounces that the electrons perform in that region (rarely more than one), in contrast to simulations with stronger $\Phi_{\parallel}$.\cite{Egedal2009}

It might seem that the counterstreaming beams of populations I and II can be at the same physical location only due to the returning electrons through the periodic $y$ boundaries (since they are ejected through different separatrices from the X-line). But the beams with $-v_z$ have also a significant contribution from population III with electrons changing the direction of their $v_z$ speed, allowing a direct interaction between counterstreaming beams. We can make a crude estimation by detecting the number of traced electrons changing direction from $+v_z$ to $-v_z$ in the exhaust region, keeping the negative drift speed at the end of the considered time period. They are 2083, but many of them with low $v_z$ should be discarded since they do not contribute significantly to the beam. By establishing a bottom threshold of minimum change in $\Delta v_z/v_{th,e}>1.0$, we get 1741, and  for $\Delta v_z/v_{th,e}>2.0$ we get 852. The beam population with $-v_z$ has 5794 traced electrons (while the beam population with  $+v_z$ has 6801). Therefore, the contribution of the population III to the counterstreaming beam $-v_z$  can be estimated to be between $(852/5794)\cdot100\%\approx15\%$ and  $(1741/5794)\cdot100\%\approx 30\%$. The rest should come from the returning electrons through the periodic boundaries. Therefore, a significant proportion of the instability and turbulence reported for the $b_g=3$ case should always be seen, independent of any boundary condition.

\subsection{Beam instabilities}\label{sec:beaminstabilities}
\begin{figure*}[!ht]
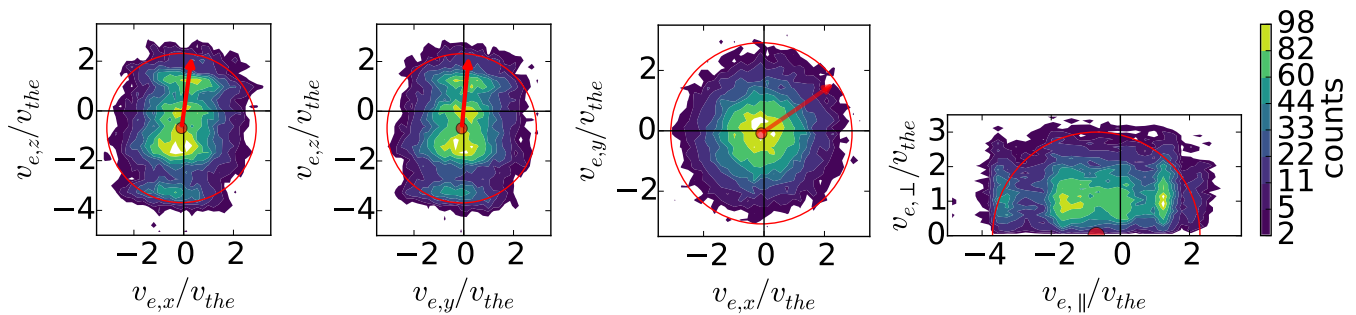
\centering
	\includegraphics[width=0.99\linewidth]{{{./detailfullvelspace_bg3_later2}}}
	\caption{Full EVDF in the strong guide-field case $b_g=3$ at $t=20\Omega_{ci}^{-1}$ (compare with the EVDF for earlier time shown in Fig.~\ref{fig:vdfs_bg}(c)). This EVDF was taken at the same location shown in Fig.~\ref{fig:jz_bg}(c). \label{fig:vdf_bg3_later}}
\end{figure*}

For moderately strong guide fields ($1.5\lesssim b_g\lesssim 5$), fast beams are accelerated causing turbulence due to streaming instabilities.
As a result, structures resembling phase space holes are formed while pitch-angle scattering isotropize the originally anisotropic beam EVDFs.  The beam EVDFs (c.f. Fig.~\ref{fig:vdfs_bg}(c)) are, therefore, transient in their interaction with turbulence. Plateaus formed in the EVDFs fill the voids in the velocity-space between the beams, making them stable (see Fig.~\ref{fig:vdf_bg3_later}). Note that these non-ideal effects and instabilities in the exhaust and separatrices region of reconnection might become weaker for higher (more realistic) mass ratios $m_i/m_e$, as shown in some previous works.\cite{Egedal2015}

\begin{figure*}[!ht]
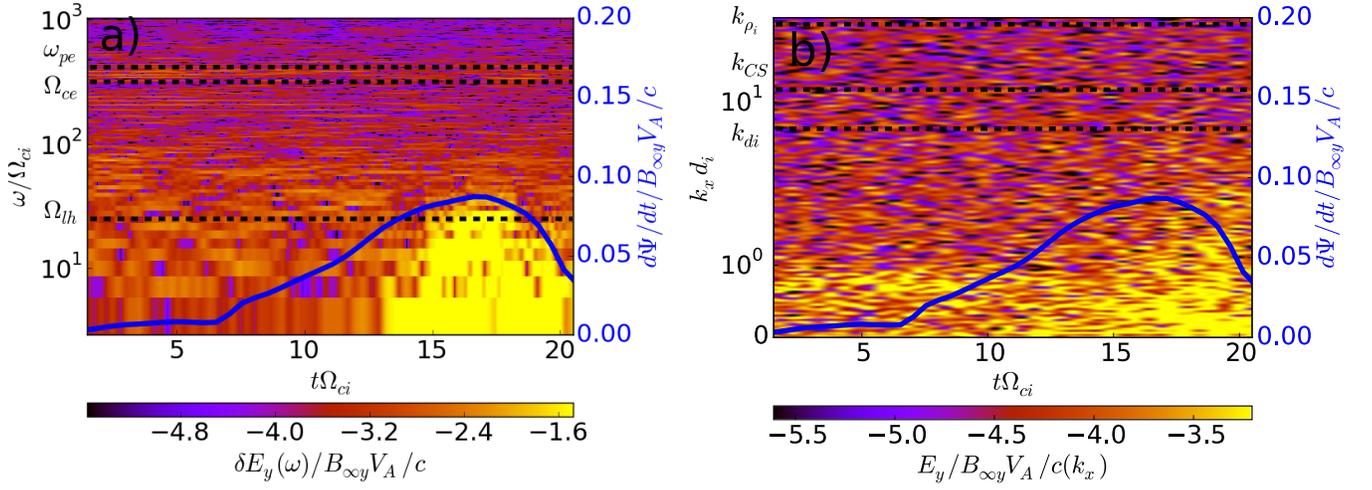
\centering
	\includegraphics[width=0.99\linewidth]{{{./spectrograms_stack}}}
	\caption{a): Spectrogram of the electric  field component $E_y$ for the case $b_g=3$, at the same location used to obtain the EVDF in  Fig.~\ref{fig:jz_bg}(c). b) Stack plot of Fourier modes $E_y(k_x)$ of an $x-$cut at constant $y=2.5d_i$, shown as dashed lines in Fig.~\ref{fig:jz_bg}(c). The solid blue line and corresponding right axis label in both plots show the time history of the normalized reconnection rate $d\Psi/dt$, where $\Psi$ is the difference in the vector potential $A_z$ between the X and O point.
		\label{fig:probe_info}}
\end{figure*}

The strong electrostatic turbulence start after $ t\gtrsim 15 \Omega_{ci}^{-1}$ (and close to the reconnection rate peak time), as one can see in the $\delta E_y$ electric field spectrogram shown in Fig.~\ref{fig:probe_info}(a). The magnetic fluctuation strength is lower than the electric one, as observed in the PSBL of the Earth's magnetosphere under low plasma-$\beta$ conditions.\cite{Zhou2014c}

Note that the electrostatic fluctuations reported here seem different from the parallel (and bipolar) turbulence usually seen in the ``low-density  separatrices'' (cavities) of guide-field reconnection due to streaming instabilities.  The main reason is that they have mostly low frequencies (see  discussion later) and are not constrained to the separatrix region, but fill most of the exhaust region.
The instabilities generating electrostatic turbulence in the ``low-density  separatrices'' of reconnection were found  to be mostly of streaming-type, generating mostly high frequency waves at electron scales.
In particular, two-streaming and Buneman instabilities seem to be very ubiquitous, as shown by 2.5D PiC simulations of magnetic reconnection with\cite{Goldman2008} and without guide-field.\cite{Fujimoto2006a,Fujimoto2011b,Fujimoto2014} Shear flow instabilities like  Kelvin-Helmholtz instability have also been reported\cite{Pritchett2009,Divin2012}(see more details in Chp.~8.4.1 of Ref.~\onlinecite{Gonzalez2016a} and references therein). Streaming instabilities due to electron beams have been observed in the PSBL of the Earth's magnetosphere, generating high frequency Langmuir-like waves (see, e.g., \onlinecite{Parks1984,Omura1996,Hwang2014} and references therein), as well as in the magnetopause.\cite{Ergun2016}

We found evidence that the instability operating in our case might be due to a combination of Buneman (electron beams with ions) and two-streaming (between electron beams) instability with parallel propagation (see a general overview of these instabilities in Ref.~\onlinecite{Omura2003} and Sec.~3.2 and 8.5 of Ref.~\onlinecite{Gary1993}) in addition to a (beam-driven) lower-hybrid instability with nearly perpendicular propagation,\cite{Coppi1976, Papadopoulos1976,Migliuolo1985,McMillan2006} with the latter being probably the dominant one. Note that most of the relative drift speed between the beams is in the out-of-plane $z$ direction because of the strong magnetization for the guide field $b_g=3$. However, there is a small in-plane projection of that relative drift that allows the existence of parallel propagating waves and the mechanism behind all these instabilities in the reduced 2.5D geometry of our simulations. The main result of these instabilities is the generation of lower hybrid waves and turbulence. An alternative way of interpreting this is considering the in-plane drift speed of the electron beams as a transverse current perpendicular to the local magnetic field (mostly along $z$), generating a cross-streaming instability known as modified two-streaming instability MTSI  (see e.g., Refs.~\onlinecite{McBride1972,Wu1983} and Chp.~4.4 in Ref.~\onlinecite{Treumann2001a}). MTSI was proposed to be active in the diffusion region of magnetic reconnection.\cite{Fujimoto2003} Note that this can only be an effect of our simulation geometry. In any case, the features of this instability are very similar to those of the lower-hybrid instability, leading to the same kind of unstable waves and turbulence.
In addition, other instabilities might also be active, like electron whistler, due to the initial anisotropy of the beams (see Sec.~7.3.2-3 in Ref.~\onlinecite{Gary1993}).

We found four signatures of these streaming instabilities generating lower hybrid turbulence. First: the characteristic frequency of the electric field fluctuations  is broadband with (low) frequencies up to the lower hybrid frequency $\Omega_{LH}=\omega_{pi}/\sqrt{1+\omega_{pe}^2/\Omega_{ce}^2}$ (See Fig.~\ref{fig:probe_info}(a)). For our parameters, this corresponds to $\Omega_{LH}=25\Omega_{ci}$.
The most likely source of free energy for this wave activity is the lower-hybrid instability due to parallel propagating electron beams (see Fig.~\ref{fig:vdfs_bg}(c)), possibly interacting also with the ion distribution (practically at $v_{\parallel}=0$), agreeing in order of magnitude with theoretical predictions.\cite{Coppi1976, Papadopoulos1976,Migliuolo1985,McMillan2006} Note that monochromatic wave activity,  such as typically expected only from two-streaming/Buneman instabilities, generates electron trapping in coherent structures, with high frequency waves close to electron scales, not observed in our case. On the other hand, the obliquely propagating lower hybrid waves are incoherent, broadband, and able to scatter particles,\cite{McMillan2006} which agree more with our simulations results. Furthermore, the typical decay period of the electron beams is within the order of magnitude of the growth rate of these instabilities (for similar parameters, Ref.~\onlinecite{McMillan2007} reported flattening of EVDFs beams in $70\Omega_{LH}^{-1}$ in their PiC simulations, which is about $3\Omega_{ci}^{-1}$ in our case).

Note that 3D PiC simulations of guide-field reconnection have shown that waves in the lower hybrid range of frequencies can be generated by either the lower hybrid drift instability (LHDI) due to density gradients\cite{Scholer2003}, or beam-driven lower hybrid instability caused by the nonlinear decay  of  two-streaming/Buneman instabilities in the cavities of the ``low-density separatrix''.\cite{Drake2003,Che2009,Che2010}
2.5D simulations of magnetic reconnection without guide field have also found wave activity close to the lower hybrid frequency generated by the Buneman instability,\cite{Fujimoto2014} in scenarios where the relative streaming speed is relatively slow. This interplay between the nonlinear phase of the structures generated by the Buneman instability and the subsequent generation of lower hybrid waves have also been reported in simulations specifically designed to study these instabilities.\cite{Reitzel1998,Miyake2000}

The second evidence for this streaming instability is its threshold or stabilization condition. For the following estimations, we should estimate first the typical relative drift speed between the electron beams. The field aligned speed between the two counterstreaming beams in  Fig.~\ref{fig:vdfs_bg}(c): $V_{{\rm rel-beams},\parallel}/v_{th,e}=2.0-(-3.5)\approx 5.5$, while the local magnetic field direction is given by the unitary vector $\hat{b}=0.1647\hat{x} + 0.1129\hat{y} + 0.9789\hat{z}$. Then, the magnitude of the projection of the field aligned speed onto the plane $x$--$y$ can be obtained by multiplying $V_{{\rm rel-beams},\parallel}$ by the factor $\sqrt{\hat{b}_x^2+\hat{b}_y^2}\approx 0.2$, resulting in $V_{{\rm rel-beams},\parallel,xy}\approx 11v_{th,i}$ (note that $v_{th,e}=10v_{th,i}$). The relative drift speed with respect to the practically unmagnetized ions is about half of these values. This has to be compared with the parallel beam temperature, given by $v_{th,e,{\rm beam},\parallel}/v_{th,e}\approx0.58$. The corresponding projection is $v_{th,e,{\rm beam},\parallel,xy}/v_{th,i}\approx1.1$. Both temperatures (field aligned or in-plane projection) are much smaller than the relative electron beam drift speed, satisfying the criterion to trigger the beam-driven lower hybrid instability.

The lower hybrid instability is stabilized for high plasma-$\beta$ conditions, because it requires negligible magnetic shear.\cite{Migliuolo1985,McMillan2006} The instability is allowed only for drift speeds $V_d$ such as $\beta_i V_d/(2v_{th,i})\ll 1$. For an in-plane drift speed of  $11v_{th,i}$, this ratio is $0.27$, which is therefore well satisfied. For smaller guide fields, even if the same kind of electron beams would appear (which is not the case), they would be stable since the higher plasma-$\beta$ would violate the previous condition. Quasilinear\cite{McMillan2006} and kinetic PiC simulations\cite{McMillan2007} have shown that the lower hybrid instability can be active in a wide range of parameters in which other parallel propagating instabilities (such as Buneman) are stable. This might be the case when the EVDFs have plateau-like structures between the beams, as it is in our case at some locations.

The third evidence is that wave activity in the lower-hybrid frequency range causes parallel electron heating and perpendicular ion heating  (see, e.g., Refs.~\onlinecite{McMillan2006,McMillan2007}, and references therein). This is because the lower hybrid waves have long parallel wavelengths (resonating with electrons) and short perpendicular wavelengths  (resonating with ions). Indeed, lower hybrid waves are known to propagate nearly perpendicularly with $k_{\parallel}/k_{\perp}\approx \sqrt{m_e/m_i}$ (see some additional features and approximations of the dispersion relation of these waves in Ref.~\onlinecite{Verdon2009}). Both features are recovered in our simulations results (the ion distributions are not shown here). This characteristic heating due to lower hybrid waves was also observed in the PSBL of the Earth's magnetosphere,\cite{Wygant2002} although due to kinetic Alfv\'en waves (KAWs).
The fourth evidence for the streaming instability is its phase speed and wave number. For that, Fig.~\ref{fig:probe_info}(b) shows the wavenumbers $k_x$ of the electric field $\delta E_y$ fluctuations. The  wavenumbers with higher spectral power show a broadband distribution (as expected from lower hybrid activity\cite{McMillan2006}) in the range $0.3\lesssim kd_i\lesssim 4$. The low end range can be misleading because the homogeneity assumption of linear theory breaks down at such large spatial scales (closer to the CS halfwidth). This wavenumber ($k_x$) contains both parallel and perpendicular fluctuations, but we can take it as $k_{\parallel}$ for an estimation on the order of magnitude. Therefore,
the range of wave phase speeds is on the order of $V_d/v_{th,i}=\omega/k_{\parallel}/v_{th,i}=6-81$ for $\omega\sim \Omega_{LH}/2$. This implies that at least the waves within the lower phase speed range can resonate with the (in-plane projected) electron beams. Note that the possibly also active  anisotropy-driven electron-whistler instability perhaps also contributes to the turbulence, as one can see at smaller scales than the size of the phase space structures (in the upper range of unstable $k$).

Note that these electrostatic structures seem different from the known electron holes   located in the ``low-density separatrix'' region of guide-field reconnection. The latter are as large as eight times the electron skin depth $d_e$, decreasing in units of $d_i$ for higher mass ratios, and mostly independent on the guide-field and mass ratio\cite{Lapenta2010} (see also Chp.~8.4.1 of Ref.~\onlinecite{Gonzalez2016a}). The size of the electrostatic structures in the exhaust region of reconnection discussed here is, however, inversely proportional to the guide-field. Their sizes coincide with the electron holes previously reported only in the order of magnitude for the case $b_g=3$.

Note that several other effects can hinder the applicability of linear theory to our case. This requires magnetized electrons with $k_{\perp}\rho_e<1$, relatively well satisfied in our case. But the approximation of homogeneous plasma, the neglect of electromagnetic effects, wave-wave coupling and the relative drift speed are either not well satisfied or vary in a wide parameter range. That is why all the previous estimations should be taken with caution.

\section{Summary and conclusions}\label{sec:conclusion}

Our goal was to obtain the characteristic electron velocity space distribution functions (EVDFs) as formed by non-antiparallel (component-, finite guide-field-) magnetic reconnection in collisionless plasmas in dependence on the guide field strength/the shear angle of the magnetic field. This is to provide observable signatures and indicators for the reconnection process in space and the laboratory. Taking into account what has been already found in the past, we concentrated on the transition to larger guide fields which magnetize the electrons throughout. We also focused our presentation and discussions on the EVDFs near the separatrices and in the plasma exhaust (outflow) of reconnection.
Utilizing a 2.5-dimensional PiC code, we also self-consistently took into account the non-linear feedback of the self-generated turbulence to the re-formation of the non-Maxwellian EVDFs.

Strong guide fields ($1.5\lesssim b_g \lesssim 6$) correspond to a small plasma-$\beta_{e,\infty}$. This controls the electron energization. In this case, reconnection accelerates electrons into beams not only along the separatrices but also in the reconnection exhaust (outflow) region. The beams propagate preferentially in the direction of the local magnetic field, practically the guide-field direction. Their EVDFs are mostly gyrotropic and anisotropic with a higher temperature perpendicular rather than parallel to the local magnetic field ($T_{e,\parallel} > T_{e,\perp}$).

In the limit of strong guide fields, the effects of the acceleration potential $\Phi_{\parallel}$ are relatively small compared to that of the reconnection electric field. Hence, in this limit, the beam electrons are accelerated mostly by the (perpendicular, inductive) reconnection electric field rather than by parallel electric fields  in cavities near the ``low-density separatrices''. After spending longer times near the X-line, electrons are ejected almost parallel to the magnetic field into the beam population. Depending on their incidence angle and speed they end up in the ``high-'' or in the ``low-density separatrix'' regions.
Part of the accelerated electrons becomes magnetically trapped in the exhaust region. In the case of strong guide fields, a smaller number of electrons are electrically trapped as required for the model of Refs.~\onlinecite{Egedal2008,Egedal2013}. This is because the electrons lose their adiabaticity after interacting with the instabilities self-generated by the beams. By this way, one of the assumptions of the electron trapping model is not satisfied anymore.

As soon as the relative drift speed of the beams exceeds the threshold of a streaming instability, a strong  broadband electrostatic turbulence is generated in the lower hybrid frequency range. We found phase speeds and wavenumbers of the excited waves that match well the theoretical predictions of lower-hybrid streaming instabilities.
The lower-hybrid turbulence leads to pitch-angle scattering of the beam electrons. This, with the elapsing time, isotropizes the beam-EVDFs.
There is a second source of free energy, the gaps between the counterstreaming beams in the velocity space. They vanish in the interaction with the self-generated turbulence and plateaus are formed in the EVDFs which are associated with phase space holes.
Note that the beam instabilities both near the separatrices and in the exhaust are weaker the higher (more realistic) the mass ratio $m_i/m_e$ is.\cite{Egedal2015}

In the case of smaller guide-fields  ($b_g\lesssim 1.5$), the electron acceleration parallel to the X-line of reconnection is less efficient since the time the particles spend near the X-line is short and magnetic trapping is less efficient. Hence, weak guide-field component reconnection does not generate the beams discussed above. Instead, a larger part of the energy released by reconnection accelerates the bulk plasma flow in the reconnection exhaust and causes anisotropically, preferentially parallel, heated EVDFs.

In the limit of very strong guide fields ($b_g > 6$), the guide field $B_g=B_z$  is much larger than the maximum in-plane (reconnecting) magnetic field $B_{\infty y}$, and the electron beam velocity is even larger than for $ b_g \lesssim 6$.  In 2.5D simulations, however, the unstable waves are driven by the in-plane projection of the beam velocity. Due to the dominance of the guide magnetic field, this velocity projection is smaller than in the guide-field regime $b_g \lesssim 6$. That is why the turbulence generated in the case of $b_g\gtrsim6$ is weaker than that one obtains for  $1.5\lesssim b_g \lesssim 6$. As a result of the 2.5D approach, the beam energy cannot be dissipated by waves and turbulence propagating in the out-of-plane direction. We already verified that in a fully 3D simulation (not shown here), plasma waves excited and propagating in the out-of-the-reconnection-plane direction, indeed, open an additional channel for wave-particle energy transfer. The fully 3D turbulence spectrum extends well beyond the lower hybrid frequency towards higher (electron) frequencies.

Note that the formation of counter-streaming electron beams and the resulting turbulence in the reconnection exhaust is not just an artifact of the periodic boundary conditions.
Instead, a significant contribution of about $15\%-30\%$  of the counterstreaming electrons is mirrored back staying well inside the exhaust plasma outflow region, not reaching the edges of the simulation box. We confirmed this by running simulations with larger boxes in the exhaust/outflow direction, exhibiting the same beam instabilities (and a somewhat stronger turbulence, results not shown here). Note that the choice of periodic boundary conditions in the direction of the reconnection outflow describes the acceleration in magnetic islands between multiple X-lines including finite guide-field effects.\cite{Drake2006} Such structures are formed by cascading reconnection\cite{Barta2011a,Barta2011b,Zhou2015c} and as a consequence of plasmoid-unstable current sheets.\cite{Loureiro2007, Bhattacharjee2009, Loureiro2016}

In this paper, we summarized our findings about the formation of EVDFs in non-antiparallel (magnetic-field) reconnection but through symmetric (density) current sheets. Some of our results, therefore, do not apply to asymmetric-density current sheets like at the Earth's magnetopause, which we and other authors have addressed previously.\cite{Silin2006,Pritchett2008,Hesse2014} As far as the EVDF formation is concerned, the difference between symmetric and asymmetry current-sheet reconnection is the more clear-cut visibility of the crescent-shape of the distributions. The reason is the difference between the number of source electrons on the two sides of the current sheet and the resulting  additional accelerating electrostatic field generated by the pressure gradients across the current sheet.\cite{Bessho2016,Chen2016i,Shay2016,Hesse2016}

Meanwhile, the MMS mission scientists have obtained the EVDFs of finite guide-field reconnection near the magnetopause like for a case of $b_g\sim 1$\cite{Burch2016a},  a case of $b_g=2$\cite{Oieroset2016} and one of $b_g=4$.\cite{Eriksson2016d} Close to the reconnection X-line, they obtained a $T_{e, \parallel} > T_{e,\perp}$-anisotropy of the EVDFs well in accordance with our symmetric-current-sheet simulation results (detailed EVDFs not shown here).
In the next task, MMS will investigate magnetic reconnection in the Earth's magnetotail, where the reconnecting current sheets are symmetric. With regard to the strength of the reconnection guide-field, smaller ($b_g\lesssim1$)\cite{Nakamura2008a} and larger $b_g$-fields\cite{Rong2012} can be expected. If our predictions about the guide field dependence are confirmed, we can extrapolate them with more confidence to the Solar atmospheric and other astrophysical and laboratory plasmas in which the guide fields can be much larger.
Note that, for mass ratios larger than those used in this study ($m_i/m_e=100$), one can find EVDFs like the ones discussed here already at smaller guide-fields,\cite{Lapenta2011a,Ng2012} while the instabilities and associated non-ideal effects can become somehow less expressed.\cite{Egedal2015}

Let us  summarize our main predictions for non-Maxwellian EVDF-signatures in dependence on the guide-field strength:

\begin{itemize}
	\item \textbf{Very small guide fields; $\boldsymbol{b_g\lesssim 0.13}$ (or $\boldsymbol{165^{\circ}\lesssim \phi \lesssim 180^{\circ}}$)}: \\
	Already for very small guide fields, most of the non-Maxwellian EVDFs typical for antiparallel reconnection (triangular-like, crescent, rings, swirls, arc-shaped) disappear. Near the X-lines, the electrons are not distinguishable anisotropically heated and accelerated.

	\item \textbf{Moderately small guide fields;  $\boldsymbol{0.13\lesssim b_g\lesssim 1.5}$ (or $\boldsymbol{65^{\circ}\lesssim \phi \lesssim 165^{\circ}}$)}: \\
	In this guide-field range, the electrons are anisotropically heated near the X-line of reconnection with $T_{e, \parallel} > T_{e,\perp}$. The EVDFs are characterized by plateaus in the direction parallel to the magnetic field. The electrons flow mainly in the direction opposite to the local magnetic field.  Their mean bulk velocity  increases with the guide field strength, while $T_{e, \parallel}$ reaches a maximum for $b_g=0.5-1.5$. In this regime, the electron trapping model of Refs.~\onlinecite{Egedal2008,Egedal2013} applies very well.

	Near the separatrices, the shape of the EVDFs exhibits anisotropic plateaus. Inside the exhaust region of reconnection, however, the electrons are heated nearly isotropically and the EVDFs are almost Maxwellian (see Fig.~\ref{fig:phase_space}(a)). On the other hand, the electron beam drift speed is reduced near the separatrices, while it increases towards the current sheet midplane. Therefore, the combination of all of those EVDFs features are seen as crescent shapes in the phase space $x$--$v_z$ distribution (i.e., in the velocity space plane through the guide field and the direction perpendicular to the current sheet midplane).

	For this case of $0.13\lesssim b_g\lesssim 1.5$,  the anisotropic plateaus of the EVDFs in the exhaust region differ from those near the X-line: they develop just a small ``bump'' in one end of the plateau which for stronger guide fields ($b_g\gtrsim 1.5$) splits off from the EVDFs ``core'', forming two distinguishable beams.
	We estimated the maximum beam drift by the criterion of reaching a ``parallel thermal spread'' $V_{max}=V_{\parallel}-3v_{th,e\parallel}$, i.e., of reaching the value of three standard deviations from the mean.
	Here, $V_{\parallel}$ denotes the parallel bulk drift speed of the entire EVDF. This speed is in the plateau region of the EVDF close to the X-line, and somewhere between the core and the bump  of the EVDF in the exhaust region.  This speed  can be used as a rough measure of the efficiency of the acceleration process. Altogether, the thermal spread ranges from $V_{max}\sim-3.67v_{th,e}$ (for $b_g=0.13$) up to $V_{max}\sim-6.18v_{th,e}$ (for $b_g=1.5$). From simulations for a number of different guide-field strengths (results not shown here in detail), we obtained an exponential regression fitting for the thermal spread:  $V_{max}/v_{th,e} \sim C_1\exp(-C_2\cdot b_g) + C_3$, with $C_1=-4.15$, $C_2=-4.11$ and $C_3=-6.37$.  The factor $C_3$  increases for smaller asymptotic plasma$-\beta_e$ and larger temperature ratios $T_i/T_e$, while $C_1$ depends on the critical magnetization of the electrons in the guide-field ($b_{g,{\rm crit}}=0.26$, see Eq.~\eqref{eq:bg_critical}). It becomes smaller (less negative) for smaller critical guide fields (up to $C_1\approx-1.58$).
	
	\item \textbf{Large guide fields: $\boldsymbol{1.5\lesssim b_g\lesssim 6}$ (or $\boldsymbol{19^{\circ}\lesssim \phi\lesssim 67^{\circ}}$)}:  \\
	In the case of large guide-fields, the EVDFs near the X-line of reconnection are gyrotropic, anisotropically heated preferentially in the direction parallel to the magnetic field, as in the case $b_g\lesssim1.5$.  On the other hand, the parallel beam drift speed of the electrons $V_{e,\parallel}$ in this regime is larger than  the one obtained by reconnection in smaller guide-fields. The parallel temperature $T_{e\parallel}$ reaches a maximum for guide-fields close to $b_g=3$. It does not increase further with guide-fields increasing beyond  $b_g=3$. The maximum parallel thermal spread of the EVDFs is roughly the same as reached already in the case of a $b_g\sim 1.5$: $V_{max}\sim-6.18v_{th,e}$. It further only weakly depends on the guide-field strength.

	In this guide-field range, the EVDFs in the exhaust region are complex and intermittent. At maximum of reconnection, beams are formed in the whole exhaust region (see Fig.~\ref{fig:vdfs_bg}(c)). The larger $b_g$ is, the more anisotropic the beam EVDFs are. Due to the conservation of the magnetic moment (see Sec.~\ref{sec:beaminstabilities}), the near-separatrix electron beams propagate either slightly faster or with the velocity $V_{max}$ of the electrons accelerated close to the X-line. The beam electrons gain, therefore, a higher total energy.

	Note that while the relative drift speed of the beams  does not significantly depend on the guide-field strength, the velocity-space gap between the beams is deeper for stronger $b_g$. At time scales of the order of $~\Omega_{LH}^{-1}$, the EVDFs exhibit phase space holes and stable plateaus while the anisotropy of the beam distribution is reduced.

	\item \textbf{Very large guide fields: $\boldsymbol{b_g\gtrsim 6}$ (or $\boldsymbol{\phi\lesssim 19^{\circ}}$)}: \\
	In case of very large guide fields, the EVDFs are similar to those formed in the large guide-field regime ($1.5\lesssim b_g\lesssim 6$). The beams do not generate, however, a significant turbulence. Hence, the lifetime of the beams is extended and becomes larger than that of the EVDF phase space holes and plateaus. In this limit, the energy of fully 3D beams is dissipated by turbulence with characteristic wave vectors pointing in the guide field direction (preliminary results, not shown here).
\end{itemize}

\begin{acknowledgments}
	We acknowledge the developers of the ACRONYM code (Verein zur F\"orderung kinetischer Plasmasimulationen e.V.).
	In particular, we are most grateful to Patrick Kilian for his helpful discussions and valuable comments.
	P.M. acknowledges funding by the Max-Planck/Princeton Center for
	Plasma Physics.

	All authors thank the referee for valuable and constructive comments which helped us to improve the discussion of our results in a wider context.
\end{acknowledgments}

\end{document}